\documentclass[sigplan,10pt]{acmart}
\renewcommand\footnotetextcopyrightpermission[1]{}
\usepackage{blindtext}

\usepackage{tikz}
\usepackage{amsmath}

\usepackage{algorithm}
\usepackage{algpseudocode}

\usepackage{xspace}

\usepackage{tabularx}
\usepackage{pifont}
\newcommand{\cmark}{\ding{51}} 
\newcommand{\xmark}{\ding{55}} 
\usepackage{array}
\usepackage{multirow} 
\usepackage{lipsum}
\usepackage{url}
\usepackage{graphicx}

\usepackage{subfigure}
\usepackage{makecell}
\usepackage{caption}
\AtBeginDocument{%
  }

\newcommand{\fref}[1]{\mbox{Fig.~\ref{#1}}}

\newcommand{\mypara}[1]{\smallskip\noindent\emph{#1}\xspace}
\newcommand{\myparab}[1]{\vspace{0.025in}\noindent\textbf{#1}}

\newcommand{\ie}{{\it i.e.}}

\newcommand{\eat}[1]{}

\newcommand{\system}{{$Xsim$}\xspace}

\newcommand{\systemAWG}{{$SUTRA_{AWG}$}\xspace}

\usepackage{color}
\newcommand{\rc}[1]{{\color{red}{#1}}}
\newcommand{\bc}[1]{{\color{blue}{#1}}}
\usepackage{graphicx}
\usepackage{makecell}
\usepackage{caption} 
\usepackage{titlesec}
\renewcommand\footnotetextcopyrightpermission[1]{} 
\setcopyright{none}

\acmDOI{}

\acmISBN{}

\acmPrice{}

\begin{document}
\title{Simulating Unified Tensor Resharding in
heterogeneous AI systems}



\author{Sumit Kumar\textsuperscript{1},
Sayantan Dasgupta\textsuperscript{1},
Kushal Mitra\textsuperscript{1},
Meet Dadhania\textsuperscript{2},
Rohan Sudhir Basugade\textsuperscript{1},
Praveen Tammana\textsuperscript{2},
Satananda Burla\textsuperscript{3},
Abed Mohammad Kamaluddin\textsuperscript{4},
Rinku Shah\textsuperscript{1}}

\affiliation{%
  \textsuperscript{1}IIIT-Delhi \country{India}  \hspace{1em} 
  \textsuperscript{2}IIT Hyderabad \country{India} 
  \hspace{1em} %
  \textsuperscript{3}Marvell Technology Inc. \country{USA}
  \hspace{1em} %
  \textsuperscript{4}Marvell Technology Inc. \country{India} 
}

\renewcommand{\shortauthors}{X.et al.}

\begin{abstract}

State-of-the-art AI training simulators assume homogeneous compute and network infrastructure. However, real-world training infrastructure is becoming increasingly heterogeneous since: (a) Model architectures such as multimodal and MoE exploit heterogeneity to improve device utilization, (b) Public cloud platforms often provide limited availability of homogeneous hardware due to fast hardware evolution, and (c) Large enterprises frequently deploy geographically distributed infrastructure that is both diverse and heterogeneous. In this paper, we present \system, a heterogeneity-aware simulator for distributed LLM training. \system supports: (i) Load balancing through non-uniform workload partitioning across heterogeneous device groups, (ii) Heterogeneity-aware collective communication via customized ring construction and chunk partitioning, (iii) Reusable heterogeneity-aware abstractions for emerging pipeline-parallel algorithms and non-uniform tensor resharding technique, (iv) Flexible input abstractions for specifying deployment plans with custom device groups and custom device-to-parallelism mappings, and (v) Pluggable integration with NS-3 and htsim, allowing users to trade off simulation fidelity for performance and scalability. 
Our evaluation demonstrates that \system accurately predicts training time for real-world heterogeneous deployments, with an error of less than 5\% across most heterogeneous data-parallel/tensor-parallel configurations and around 2\% error with pipeline-parallel communication modeling. We expose actionable metrics such as pipeline bubble time and straggler waiting time.

\end{abstract}

\maketitle
 \section{Introduction} 
 \label{sec:intro}

\eat{
\begin{table}[t]
\centering
\scriptsize
\caption{Co-design choices for training components in large-scale AI clusters.}
\label{tab:ai-system-design-components}
\renewcommand{\arraystretch}{1.5}
\setlength{\tabcolsep}{2 pt}
\begin{tabular}{p{3cm} p{5cm}}
\hline
\textbf{Design component} & \textbf{Design choices} \\
\hline

Parallelism strategies~\cite{duan2024efficient} &
Expert parallelism, Sequence Parallelism, Auto Parallelism, and Data/Tensor/Pipeline parallelism  \\

Transport protocols &
RoCEv2~\cite{gangidi2024rdma}, DCQCN, MLTCP, Multipathing~\cite{van2012multipathing}, UET~\cite{hoefler2025ultra}, Swift~\cite{kumar2020swift}, and Falcon~\cite{singhvi2025falcon} \\

Network topologies~\cite{duan2024efficient} &
Clos topology, BCube, Dragonfly, Rail-optimized, Rail-only \\

Interconnects &
NVLink/NVSwitch~\cite{gangidi2024rdma,tomshardware-gb300-nvl72-cluster}, InfiniBand~\cite{tomshardware-gb300-nvl72-cluster}, GPUDirect RDMA~\cite{gcp-ai-hypercomputer-networking}; RDMA Ethernet and VPC networking~\cite{alibaba-heterogeneous-computing} \\

Collective comm libraries ~\cite{duan2024efficient} &
NCCL, RCCL, HCCL, MPI, Horovod, Gloo, TPU-native collectives, and XLA-based communication  \\

Hardware configurations & GPUs, TPUs, ASICs, FPGAs, SmartNICs/DPUs, HBM, NVMe/SSD, CXL memory\\ 
\hline
\end{tabular}
\end{table}
}

Rapid evolution and scale in modern distributed training systems create a complex design space spanning parallelism strategies, transport protocols, network topologies, and hardware configurations~\cite{won2023astra}. 
To meet the workload demands at scale, leading hyperscalers tightly co-design network fabric, network topology, and accelerator interconnects for efficient distributed AI training.
For example, Meta~\cite{gangidi2024rdma} deploys dedicated RoCEv2~\cite{rocev2-fabrics} fabrics with two-stage Clos topology to provide non-blocking, low-latency collective communication~\cite{si2025collectivecommunication100kgpus} across GPU clusters. Microsoft Azure~\cite{tomshardware-gb300-nvl72-cluster} combines high-bandwidth intra-rack NVLink with InfiniBand inter-rack connectivity to scale GPU clusters.

\eat{

\myparab{Deciding appropriate compute and network infrastructure for distributed AI training is non-trivial.}
Large-scale AI training workloads impose stringent requirements on infrastructure scalability, reliability, and tail latency. To meet these demands, workload deployers carefully co-design congestion control mechanisms, network topologies, compute and network hardware, and GPU interconnects (see Table~\ref{tab:ai-system-design-components}). \bc{Add table with evolving configurations on each front.} While deployers also seek to optimize operational costs and power efficiency, these objectives must be met without sacrificing performance.
In practice, achieving this balance is difficult. For example, RDMA-based deployments depend on lossless fabrics and finely tuned congestion control mechanisms such as DCQCN~\cite{zhu2015congestion}, yet configuring and operating these mechanisms at scale remains challenging~\cite{gangidi2024rdma}.

At the same time, hardware manufacturers pursue aggressive area and resource optimizations to improve performance, reliability, and scalability. Recent proposals, such as lossy collective communication for AI training, exploit the tolerance of reductions to partial loss and packet reordering to improve tail latency and reduce NIC's memory usage~\cite{warraich2025optinic}.

}

\myparab{Understanding distributed training behavior at scale necessitates a simulator.}
The large design space induces complex interactions across workload, system, and network layers, making it difficult to isolate the impact of individual design choices in production clusters. Moreover, high-impact events such as congestion, stragglers, and tail-latency amplification are difficult to reproduce consistently in real deployments~\cite{singhvi2025falcon}.
These challenges motivate the need for a simulator that can systematically explore the design space and stress-test AI infrastructure under controlled, repeatable conditions.

State-of-the-art (SOTA) simulators such as ASTRA-sim~\cite{won2023astra}, SimAI~\cite{wang2025simai}, Multiverse~\cite{gui2025accelerating}, ATLAHS~\cite{shen2025atlahs} and Meta's Arcadia~\cite{Arcadia}, provide a full-stack simulation framework for training clusters. However, they assume homogeneous compute and networking infrastructure. This assumption does not reflect modern AI deployments (see Table~\ref{tab:hyperscale-infra-examples}). As a result, these simulators risk obscuring critical interactions, leading to misleading conclusions and suboptimal infrastructure designs~\cite{strati2025sailor,yan2024hexiscale}.

\eat{
In this paper, we present~\system, a full-stack, LLM training simulator that accurately predicts training performance for real-world AI clusters with heterogeneous compute and network infrastructure.
}

In this paper, we present~\system, a system-level LLM training simulator for real-world AI clusters with heterogeneous compute and network infrastructure.

\begin{table}[t]
\centering
\scriptsize
\caption{AI infrastructure at hyperscalers.}
\label{tab:hyperscale-infra-examples}
\renewcommand{\arraystretch}{1.5}
\setlength{\tabcolsep}{2 pt}
\begin{tabular}{p{1.2cm}  p{1.3cm} p{1.6cm} p{3.7cm}}
\hline
\textbf{Hyperscaler} & \textbf{Topology} & \textbf{Fabric} & \textbf{Compute and Interconnect} \\
\hline
Google Cloud~\cite{gcp-ai-hypercomputer-networking} &
Rail-optimized &
GPUDirect RDMA, GPUDirect TCPX. &
A4X Max (GB300 Ultra superchips), A3 Ultra (NVIDIA H200 GPUs), A3 High (NVIDIA H100 GPUs). \\

Alibaba Cloud~\cite{alibaba-heterogeneous-computing} &
SHENLONG architecture~\cite{alibaba2021shenlong} &
Hybrid RDMA and VPC networking~\cite{aws-vpc-overview} &
Heterogeneous nodes combining GPUs (L20, A10, V100, T4, P100), FPGAs, and ASICs. \\

Meta~\cite{gangidi2024rdma} &
Two-stage Clos &
RoCEv2~\cite{rocev2-fabrics} &
ZionEX with NVIDIA A100 GPUs, Grand Teton NVIDIA H100 GPUs. \\

Microsoft Azure~\cite{tomshardware-gb300-nvl72-cluster} &
Full mesh, fat tree &
RDMA~\cite{infiniBand-roce-fabrics} &
GB300 NVL72 systems. \\

\hline
\end{tabular}
\end{table}

\begin{table}[t]
  \centering
  \scriptsize
  \caption{Evolution of NVIDIA data center GPUs.} 
  \label{tab:nvidia-evolution}
  \setlength{\tabcolsep}{8pt}
  \begin{tabularx}{\linewidth}{c c c c c c}
    \toprule
    \textbf{Model} &
    \textbf{Year} &
    \textbf{Archit.} &
    \shortstack{\textbf{CUDA}\\\textbf{Cores}} &
    \shortstack{\textbf{Memory}\\\textbf{(GB)}} &
    \shortstack{\textbf{FP16}\\\textbf{(TFLOPS)}} \\
    \midrule
    A100       & 2020 & Ampere  & 6912  & 40/80  & 77.97 \\
    H100       & 2022 & Hopper  & 16896 & 80     & 204.9 \\
    H200       & 2023 & Hopper  & 16896 & 141    & 989.5 \\
    B100       & 2024 & Blackwell &~14592  & 192 & 1800 \\
    B200       & 2024 & Blackwell &~16896& 192 & 2250  \\ 
    \bottomrule
  \end{tabularx}
\end{table}

\myparab{Modern distributed training deployments are increasingly becoming heterogeneous.}
Heterogeneity in distributed AI training infrastructure arises naturally from four primary factors. 
(a) GPU compute and interconnect bandwidth advance approximately 3.0$\times$ and 1.4$\times$ annually~\cite{gholami2024ai}, respectively (see Table~\ref{tab:nvidia-evolution}). Organizations progressively integrate new GPU generations into existing clusters before retiring older hardware, resulting in a transitional heterogeneous state that may persist for months or years;
(b) Demand for larger AI models continues to surge, for example, Meta’s Llama-4 with 2 trillion parameters was trained on $32K$ GPUs \cite{llama4}. With power regulations within a given region, enterprises such as Meta run AI training across multiple data centers, each equipped with distinct hardware capabilities and interconnect architectures~\cite{meta-gigacluster}; 
(c) Since the supply of GPU resources is unable to meet the demand, some organizations rely on the cloud resources. Cloud users often cannot wait for uniform GPU allocations; instead, they accept heterogeneous configurations to meet time-to-market constraints~\cite{motivationMSR-multitenant-cloud-training,motivation-alibaba-MLaaS,mo2024heet,sagemaker}. Table~\ref{tab:hyperscale-infra-examples} shows the configurations offered by popular hyperscalers;
(d) Device heterogeneity is leveraged for better utilization. For instance, in multimodal model training, compute-intensive modalities (LLM backbone) are mapped to newer GPUs, and less compute-intensive tasks (encoder) are mapped to older hardware~\cite{zhang2025disttrain}. 

\myparab{Can we naively reuse a homogeneous deployment plan for a heterogeneous cluster?}
We observe three major roadblocks: 
(1) Equal-sized model or data partitions when mapped to devices with asymmetric compute and memory result in {\em load imbalance}, causing faster GPUs to wait for slower ones. Metis~\cite{um2024metis} reported an $8 \times$ slowdown in training throughput due to load imbalance. 
(2) Adopting unequal-sized model partitioning helps match the processing speeds. However, (a) Collective operations such as AllReduce must handle misaligned batch sizes and tensor shapes~\cite{qi2026hetauto}, and (b) Network traffic generated for gradient synchronization is deterministic but not uniform, resulting in the need for custom transport protocols to reduce training completion time;
(3) Network bottlenecks intensify as gradient and activation communications traverse hybrid interconnects.

\myparab{Motivating Example:} In our evaluation of the Llama-7B model~\cite{touvron2023llama}, SimAI~\cite{wang2025simai}, which assumes homogeneous clusters, mispredicts training time for a heterogeneous deployment by up to 80\%, whereas~\system maintains an error below 4.7\% (see~\fref{fig:qus2-plot}). This gap arises because SimAI does not capture straggler effects caused by non-uniform compute capacities during collective communication.

\eat{
\myparab{Motivation.} While reusing a homogeneous deployment plan over a heterogeneous cluster, Metis~\cite{um2024metis} reported an $8 \times$ slowdown in training throughput due to load imbalance. 
\system can be used to infer metrics such as slowdown via simulation. One of our evaluations (see Figure~\ref{fig:qus2-plot}) across Llama 7B~\cite{touvron2023llama} model shows that the SOTA training simulator, SimAI~\cite{wang2025simai}, that assumes homogeneous clusters, predicts the training time of a heterogeneous model deployment with an error rate of up to $80\%$, whereas~\system's error rate is within $4.7\%$ 
(setup details in~\S\ref{sec:eval}). 
This is because SimAI does not simulate the straggler effect due to non-uniform compute capacities during collective communication. 
}

\myparab{Resharding Under Arbitrary Heterogeneity.}
Recent work has proposed heterogeneous collective communication~\cite{heiheteccl,zhao2024forestcoll} and specialized resharding mechanisms for layout mismatches~\cite{qi2026hetauto,zhuang2023optimizing,arfeen2025nonuniform}. However, these approaches target narrow scenarios and assume fixed communication structures, limiting their applicability to modern heterogeneous training deployments that combine non-uniform data-parallel (DP), tensor-parallel (TP), and pipeline-parallel (PP) configurations. 
\system bridges this gap with a unified abstraction enabling efficient synchronization across arbitrary heterogeneous parallelism strategies.


\myparab{Evaluating training performance of heterogeneous deployment plan.}
The core idea of state-of-the-art heterogeneity-aware deployment strategies~\cite{jia2022whale,um2024metis,wu2025hetermoe, zhang2024hap,yan2024hexiscale,nie2024cannikin} is to non-uniformly partition the model across heterogeneous GPUs to improve resource utilization and performance~\cite{yan2024hexiscale}.
To evaluate the proposed deployment plan for heterogeneous clusters, researchers rely on: (a) real-world deployments~\cite{jia2022whale,um2024metis,yan2024hexiscale}, which do not have optimized collective communication implementation, and are not scalable and accessible to all, or (b) analytical simulations~\cite{strati2025sailor, tang2024simulating}, which do not mimic real-world conditions, or end-to-end full system simulation. 
These observations motivate the need for a new simulator framework that incorporates heterogeneity-aware deployment strategies and accurately models both compute and network diversity for large-scale model deployment. 

\begin{table}[t]
  \setlength{\tabcolsep}{2pt}
  \caption{Comparison of state-of-the-art distributed training }
  
  \label{tab:simulator-comparison}
  \tiny
  \begin{tabular}{|l|c|c|c|c|c|c|c|}
    \hline
    \textbf{Features} &
    \makecell{\textbf{Astra-}\\\textbf{Sim~\cite{won2023astra}}} &
    \makecell{\textbf{Echo} \\ \textbf{~\cite{feng2024echo}}} &
    \makecell{\textbf{vTrain} \\ \textbf{~\cite{bang2024vtrain}}} &
    
    \makecell{\textbf{Multi-} \\ \textbf{verse~\cite{gui2025accelerating}}} &
    \makecell{\textbf{ATLAHS} \\ \textbf{~\cite{shen2025atlahs}}} &
    \makecell{\textbf{SimAI} \\ \textbf{~\cite{wang2025simai}}} &
    \makecell{\textbf{\system} } \\
    
    \hline
    \makecell[l]{Trace extrapolation \\support} & \xmark & \cmark & \xmark  & \xmark & \xmark & \cmark & \cmark \\
    \hline
    \makecell[l]{Full stack training \\simulation} & \cmark & \cmark & \cmark & \cmark & \cmark & \cmark & \cmark \\
    \hline
    \makecell[l]{Collective optimization \\support} & \xmark & \cmark & \cmark  & \xmark & \cmark & \cmark & \cmark \\
    \hline
    \makecell[l]{Pipeline parallelism \\support (discrete-event)} & \xmark & \xmark & \xmark  & \xmark & \cmark & \xmark & \cmark \\
    \hline
    \makecell[l]{Network protocol \\simulation} & \cmark & \xmark & \xmark  & \cmark & \cmark & \cmark & \cmark \\
    \hline
    \makecell[l]{Heterogeneous cluster \\simulation} & \xmark & \xmark & \xmark & \xmark & \xmark & \xmark  & \cmark \\
    \hline
  \end{tabular}
\end{table}

\myparab{Our proposal.}
We present \system, a system-level simulator for heterogeneous AI training that models clusters with heterogeneous compute nodes\footnote{We assume homogeneous GPUs within a node, and heterogeneous GPUs across nodes and clusters to mimic real-world infrastructure.} and heterogeneous interconnects\footnote{We assume scale-up heterogeneity across nodes/clusters, whereas scale-out heterogeneity across clusters.}. \system introduces abstractions for non-uniform workload partitioning, heterogeneity-aware tensor resharding and synchronization, and pipeline-parallel execution with straggler effects. It further extends both NS-3~\cite{bai2024unison} and htsim~\cite{htsim} to model heterogeneous scale-up and scale-out networks, providing a choice between high-fidelity packet-level protocol simulation and scalable flow-level simulation. Together, these capabilities enable end-to-end simulation of heterogeneous training deployments spanning compute, communication, and network layers.

\myparab{Key Contributions.} 
To the best of our knowledge, \system is the first heterogeneity-aware, system-level simulator for AI training (see Table~\ref{tab:simulator-comparison}). 
Our key contributions include:

\noindent{(1)} We identify core requirements for heterogeneous AI training simulation (\S\ref{sec:requirements}) and design abstractions to specify non-uniform workload partitioning across heterogeneous device groups (\S\ref{subsec:input_spec}).

\noindent{(2)} 
We design Sweep-line-based DP grouping, and LCM-based multi-ring collectives and chunking algorithms to enable synchronization and resharding across mismatched DP, TP, and PP layouts (\S\ref{subsec:hetero_ring}, \S\ref{subsec:hetero_chunk_post_ring}). 
We design primitives for collective communication and tensor resharding, which we leveraged to implement two SOTA tensor resharding frameworks: HetAuto~\cite{qi2026hetauto} and AlpaComm~\cite{zhuang2023optimizing}.


\noindent{(3)} 
We extend the runtime to model heterogeneous pipeline-parallel execution (e.g., pipeline barriers, GPipe-style communication, straggler effects), providing reusable primitives for future PP strategies (\S\ref{subsec:barrier_dp_pp}).


\noindent{(4)} 
We extend NS-3 and htsim network simulators to support heterogeneous scale-up/scale-out networks, capturing variations in link bandwidth, PCIe interconnects, NIC/scale-up delays, and topologies (\S\ref{subsec:network_simulation}).


\noindent{(5)} 
We demonstrate that \system is accurate and scalable: it predicts training time within 7\% error for NVLink, predicts training time within 17\% error for hybrid (NVLink, PCIe, and Ethernet in non-isolated cluster), speeds up network simulation by up to 47$\times$ via htsim, and exposes actionable metrics (e.g., pipeline bubble time, straggler waiting time, TCO (Total Cost of Ownership)) for capacity planning (\S\ref{sec:eval}). \system's heterogeneity-aware resharding algorithm reduces synchronization overhead by up to 21\% compared to the state of the art.

\section{Background and Motivation} \label{sec:background}  

\subsection{Emergent Challenges with Heterogeneity in AI Training Clusters}
\label{subsec:hetero-train-components}
We observe heterogeneity-induced challenges in utilization, deployment complexity, and communication.

\myparab{Challenge 1: Load Balancing Across Asymmetric Hardware.}
{\em Uniform workload partitioning} is the default strategy in frameworks like Megatron-LM~\cite{shoeybi2019megatron} and DeepSpeed~\cite{rasley2020deepspeed}. Training frameworks assign equal portions of model layers, tensor slices, or data batches to all devices regardless of their compute capabilities. In heterogeneous clusters, this approach causes severe performance degradation due to compute imbalance. That is, high-performance GPUs (e.g., H100 with 204.9 FP16 TFLOPS) complete their assigned work significantly faster than lower-tier GPUs (e.g., A100 with 77.97 FP16 TFLOPS), leading to substantial idle time during synchronization barriers. 
Heterogeneity-aware training frameworks such as HexiScale~\cite{yan2024hexiscale} and Metis~\cite{um2024metis} use {\em non-uniform partitioning} strategies that assign layer/tensor/batch slices proportional to the device's compute and memory capabilities, and create custom device groups (DG) \footnote{ A {\em device group} represents a set of compute nodes with homogeneous compute and interconnect properties, modeled as a single unit for assigning parallelism strategies.}.

\myparab{How can a simulator help?} 
A heterogeneity-aware simulator can track {\em GPU idle time} during synchronization barriers, and evaluate different partitioning strategies. This enables researchers to validate their workload distributions that maximize cluster utilization before deploying to real hardware.

\myparab{Challenge 2: Synchronization Overhead and Tensor Shape Mismatches}
During gradient synchronization, asymmetric workload partitioning leads to mismatches in tensor shape and batch size. This requires additional resharding operations to restore compatibility. Collective operations, such as AllReduce or broadcast, must combine/split varying tensor shapes and batch sizes across device groups with different degrees of tensor parallelism (TP) and Data parallelism (DP). 
Heterogeneity-aware training solutions repartition and redistribute tensor data, which incurs: (a) computation overhead: to split/combine gradient tensor before participating in collective operation, and (b) communication overhead: to exchange reshaped tensor segments using custom collective operations, i.e., multigroup DP ring \footnote{A multigroup DP ring is a data-parallel communication ring that spans devices belonging to multiple device groups, enabling gradient synchronization across heterogeneous hardware groups.}, which adds network traffic beyond standard gradient synchronization.

\myparab{How can a simulator help?} 
A heterogeneity-aware simulator enables deployers to decide an optimal deployment plan after evaluating varying TP degrees, chunking strategies, and communication topologies (such as multi-ring configurations) without accessing vast, expensive physical clusters. 

\myparab{Challenge 3: Communication Bottlenecks in Heterogeneous Networks}
Modern clusters deploy heterogeneous interconnects: NVLink (Gen3: 600 GB/s vs Gen4: 900 GB/s), PCIe (Gen4: 512 Gbps vs Gen5: 1024 Gbps), and RDMA NICs (ConnectX-6: 200 Gbps vs ConnectX-7: 400 Gbps )~\cite{nvidia-infra}.
In blocking collective communication operations such as AllReduce, networks with slower links significantly increase communication tail latency, leading to a substantial increase in end-to-end training performance~\cite{yan2024hexiscale}.

\myparab{How can a simulator help?}
A heterogeneity-aware simulator with network modeling can capture asymmetric bandwidth, latency, congestion, and tail latencies across mixed interconnects. By simulating communication patterns prior to deployment, deployers can identify bottlenecks in specific heterogeneous configurations, evaluate alternative topology mappings, and predict which interconnect placements or workload partitioning strategy can minimize communication overhead.

\subsection{Heterogeneity-aware AI Training Deployment: An Example}
\label{subsec:hetero-training-example}

State-of-the-art heterogeneity-aware AI training systems jointly optimize device grouping, hybrid parallelism (PP, TP, DP), and non-uniform partitioning of data, layers, and tensors. By mapping each device group to an appropriate parallelism configuration and degree, these systems improve resource utilization and reduce end-to-end training time.

\begin{figure}[t]
    \centering
    \includegraphics[width=0.6\linewidth]{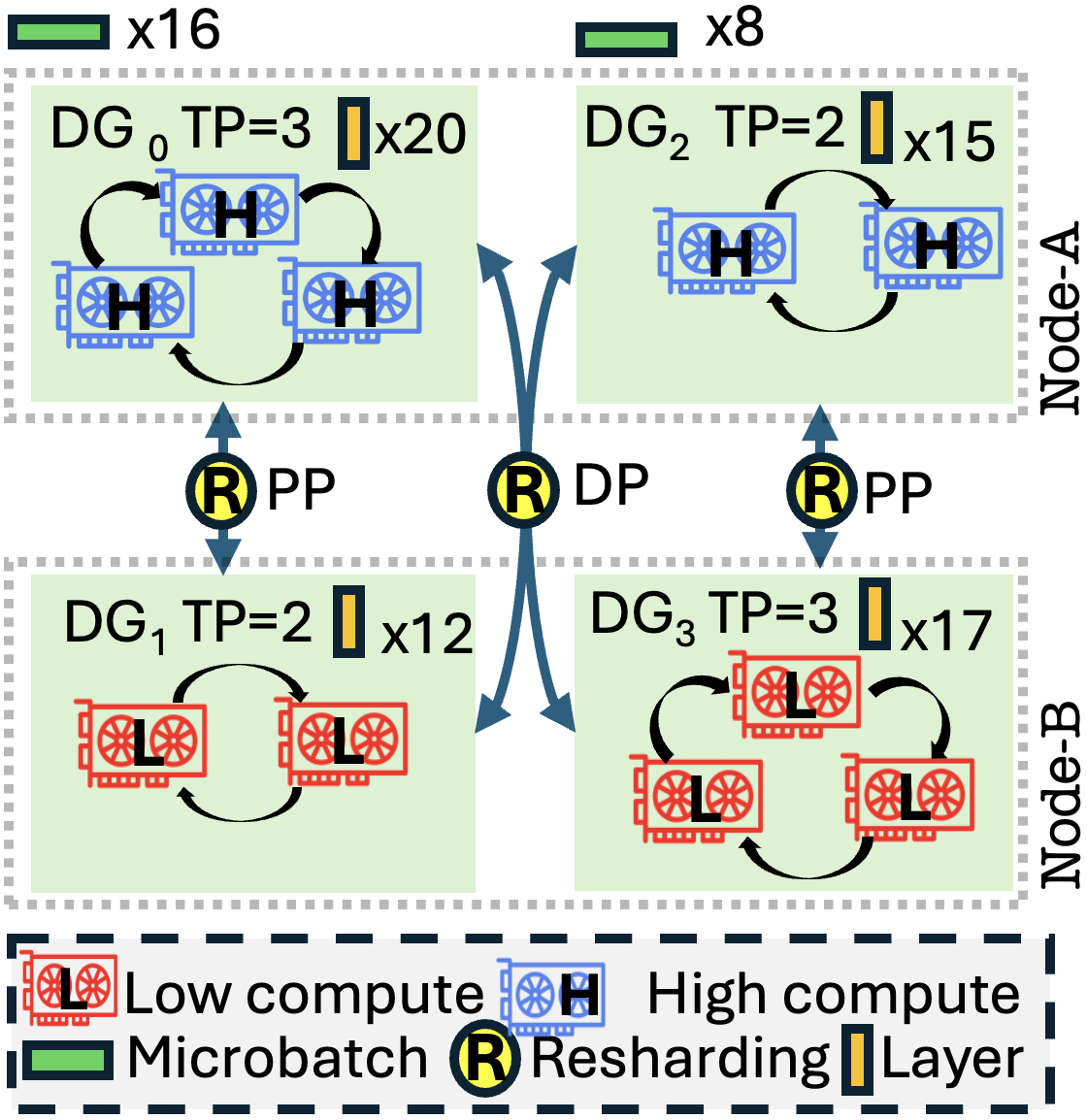}
    \caption{Example deployment with heterogeneous compute. 7B model is partitioned across two nodes, i.e., one node with 5$\times$H100 ("H") and the other with 5$\times$A100 ("L"). The model and data batches are non-uniformly partitioned to efficiently utilize resources.} 
    \label{fig:hetero-training-example}
\end{figure}

Figure~\ref{fig:hetero-training-example} illustrates a hybrid parallelism configuration for training the Llama-2 (7B)~\cite{touvron2023llama} model with $32$ layers on a heterogeneous cluster. The cluster consists of two nodes:  $Node_A$, equipped with $5 \times H100$ 80G (in blue), and $Node_B$, equipped with $5 \times A100$ 40G GPUs (in red).

\noindent{\em \textbf{Non-uniform workload partitioning.}} To effectively utilize this heterogeneous setup (\fref{fig:hetero-training-example}), the workload is non-uniformly partitioned across device groups, with different batch sizes across DP groups (i.e., 8 and 16 are marked as green), different tensor-parallel degrees (e.g., TP = 2 and TP = 3), and asymmetric pipeline partitioning (i.e., 20 layers assigned to $DG_0$ and 12 layers to $DG_1$ marked as yellow).

\noindent{\em \textbf{Resharding.}} 
During the backward pass, gradient synchronization across device groups must be preceded by resharding to ensure a common tensor shape and partitioning. Resharding is unavoidable when communicating groups use different TP degrees. Figure~\ref{fig:hetero-training-example} shows that all inter-node communication between H100 and A100 device groups involves mismatched TP degrees and therefore requires resharding before collective operations. In contrast, pipeline parallelism (PP), if implemented in isolation, does not require resharding, as inter-stage communication is point-to-point and sequential despite differing group sizes. Some heterogeneity-aware systems (e.g., HetPipe~\cite{park2020hetpipe}) avoid resharding by adopting a parameter-server–based synchronization model, at the cost of increased communication and potential server-side bottlenecks.

\subsection{State-of-the-art AI Training Simulators}
\label{subsec:sota-simulators}

State-of-the-art simulators (see Table~\ref{tab:simulator-comparison}) can be classified into two broad categories: (1) workload generators~\cite{sridharan2023chakra} and (2) full-stack training simulators~\cite{won2023astra,wang2025simai,feng2024echo,gui2025accelerating,shen2025atlahs}. 

Chakra\cite{sridharan2023chakra} executes real-world training workload on GPUs to capture the dependencies between compute, communication, and memory operations to generate a real-world workload trace. 
Further, SimAI~\cite{wang2025simai}, Echo~\cite{feng2024echo}, and Multiverse~\cite{gui2025accelerating} extrapolate the real-world traces obtained for a subset of the GPU cluster, whereas NeuSight~\cite{lee2025forecasting} and vTrain~\cite{bang2024vtrain} are  simulators that analytically predict training time. AstraSim~\cite{won2023astra}, SimAI~\cite{wang2025simai},  ATLAHS~\cite{shen2025atlahs}, and Multiverse~\cite{gui2025accelerating} simulate the entire training stack, including the workload generator, workload partitioning for parallelism, partition assignment to the set of GPUs (\ie, device groups), collective communication management, generation of compute and communication events, event scheduling, simulating compute events across GPUs, and simulating communication events over the training cluster.  

However, none of the existing works support LLM training in a heterogeneous cluster setting that comprises varied device (GPU and interconnect) capabilities and configurations. \system fills this gap.

\begin{figure}[t]
    \centering
    \includegraphics[width=0.35\textwidth]{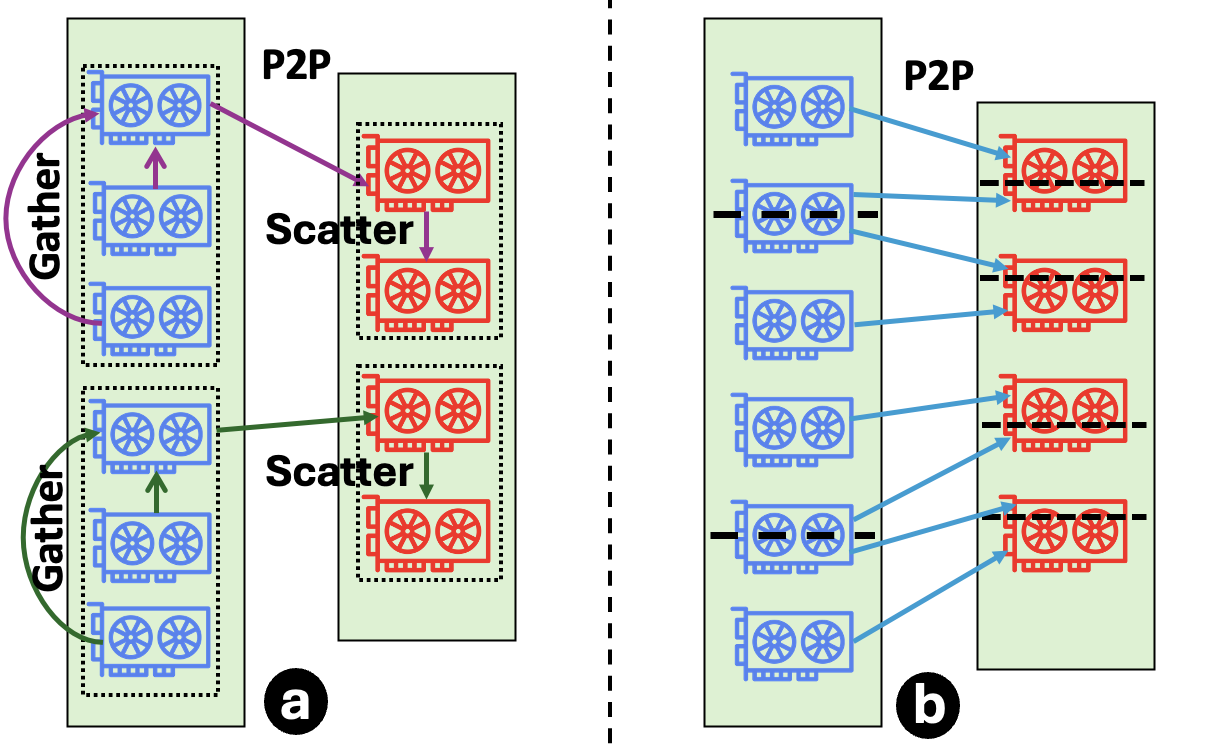}
    \caption{Comparison of SOTA tensor resharding approaches for transferring a 12-element global tensor from a source stage ($TP=6$) to a destination stage ($TP=4$). (a) HetAuto employs a 3-phase hierarchical approach (Gather $\rightarrow$ P2P $\rightarrow$ Scatter) that groups devices into two virtual clusters, each governed by $\text{GCD}(6,4)=2$, and routes traffic through leaders. (b) AlpaComm establishes direct point-to-point connections based on irregular cutpoint intervals, leading to non-uniform chunks.} 
    \label{fig:resharding-algo}
\end{figure}


\subsection{State-of-the-art tensor resharding approaches}
\label{subsec:tensor_resharding_sota}


Heterogeneous LLM training often uses different DP, TP, and PP degrees across device groups, creating tensor layout mismatches that require resharding~\cite{um2024metis,yan2024hexiscale,qi2026hetauto}. Prior work~\cite {qi2026hetauto,zhuang2023optimizing} proposes optimized resharding mechanisms for these layouts. In this section, we provide an overview of these works using an example.


\fref{fig:resharding-algo} (a) shows HetAuto~\cite{qi2026hetauto} addresses resharding across heterogeneous pipeline stages, i.e., $TP = 6 \rightarrow TP = 4$, using a GCD-based three-phase strategy. It partitions source and destination devices into virtual groups based on the GCD of their TP/DP degrees.
For example, between $TP = 6$ and $TP = 4$ stages, HetAuto uses $\mathrm{GCD}(6,4)=2$ to form two virtual groups and coordinate data movement through leader nodes.
Data is then reorganized through three phases: (i) intra-cluster gather to leader nodes, (ii) cross-cluster P2P transfer between leaders, and (iii) intra-cluster scatter to destination ranks.

To resolve tensor partitioning mismatches, AlpaComm~\cite{zhuang2023optimizing} applies a cutpoint-union algorithm that computes the union of shard boundaries from both layouts, partitioning the global tensor into atomic communication units. For example, when resharding a 12-element tensor from $TP = 6$ to $TP = 4$ (\fref{fig:resharding-algo} (b)), the source boundaries $\{0,2,4,6,8,10,12\}$ and destination boundaries $\{0,3,6,9,12\}$ produce a union boundary set of $\{0,2,3,4,6,8,9,10,12\}$. These boundaries divide the tensor into eight irregular, non-uniform communication units $ [2, 1, 1, 2, 2, 1, 1, 2]$, which are then explicitly mapped between sender and receiver ranks.

\section{Design Requirements} 
\label{sec:requirements}


Heterogeneity-aware distributed AI training solutions such as Hexiscale~\cite{yan2024hexiscale}, Metis~\cite{um2024metis}, FlashFlex~\cite{yan2024flashflex}, Cephalo~\cite{guo2024cephalo}, and Atlahs~\cite{shen2025atlahs} solve complex combinatorial optimization problems to generate a deployment for training within a heterogeneous cluster.
The deployment comprises: (a) a heterogeneous compute cluster, (b) a heterogeneous cluster and host topology for scale-out, and scale-up, respectively, and (c) asymmetric workload to device mapping. 
Among other things, a heterogeneity-aware AI training simulator must have the provision to ingest these specifications. 

To enable realistic simulation of LLM training on heterogeneous infrastructure, we have identified the key abstractions {\em "A"} and components {\em "C"} that a heterogeneity-aware full-stack simulator must expose and implement.

\myparab{[A1] Device Group and Model Specification}~(\S\ref{subsec:input_spec}). 
The simulator must support abstractions to ingest heterogeneous compute and AI model configurations. For each device group (DG), the specification must include information such as the PP stage, DP stage, number of model layers assigned, micro batch size, and the number of compute nodes (GPUs) in the device group. 

\myparab{[A2] Heterogeneous Cluster and Host Topology Specification}~(\S\ref{subsec:input_spec}).
The simulator must ingest detailed network hardware specifications to accurately model performance variance. The abstraction should include the scale-up (host) and scale-out (cluster) topologies, along with link bandwidth and delays. The deployer can simulate various interconnect types, such as NVLink, PCIe, Ethernet, and InfiniBand, using appropriate topologies and bandwidth/delay configurations.

\myparab{[C1] Asymmetric Workload Generation}~(\S\ref{subsec:workload_gen}).
Each device group profile may comprise a distinct number of model layers, tensor size, and micro-batch size. The simulator must generate a device-group-specific workload file (i.e., trace) and support Multiple-Instruction-Multiple-Data (MIMD) orchestration. 

\myparab{[C2] Heterogeneity-aware Ring Construction}~(\S\ref{subsec:hetero_ring}).
The simulator must implement an algorithm that constructs an optimal ring for gradient synchronization in a heterogeneous cluster.
The algorithm must automatically identify valid DP groups (which may overlap) and select an optimal DP communication group to form the ring.

\myparab{[C3] Heterogeneity-aware Chunk Partitioning}~(\S\ref{subsec:hetero_chunk_post_ring}).
To handle tensor shape mismatches caused by heterogeneous TP sizes, the simulator must implement an algorithm that appropriately partitions the gradients into chunks, groups them, and shares them during collective operations, such as AllReduce. 

\myparab{[C4] Pipeline-Parallel Communication and Synchronization.}~(\S\ref{subsec:barrier_dp_pp})
A heterogeneity-aware simulator must faithfully model pipeline-parallel dependencies, including pipeline barriers and inter-stage activation/gradient communication. It must also quantify straggler effects and pipeline bubbles arising from non-uniform compute capabilities, parallelism degrees, and communication paths across stages.


\myparab{[C5] Scalable Network Simulation}~(\S\ref{subsec:network_simulation}). To verify large-scale heterogeneous deployments efficiently, the simulator should integrate network simulators that support flow-level network simulation. This approach enables modeling Flow Completion Time (FCT), congestion, and tail latencies for thousands of flows without the computational overhead of packet-level simulation.

\myparab{[C6] Heterogeneous Compute Simulation}~(\S\ref{subsec:barrier_dp_pp}). The core simulation engine must capture and model execution time based on device-specific capabilities. This involves capturing and simulating per-layer computation time scaled by the specific GPU type of the assigned device group. 


\section{\system Simulator}  
\label{sec:design}

\system is a full-system, discrete-event training simulator for heterogeneous AI clusters. It extends the principles of the state-of-the-art simulator, SimAI~\cite{wang2025simai}, which is designed for training in homogeneous clusters. 

\begin{figure}[t]
    \centering
    \includegraphics[width=0.5\textwidth]{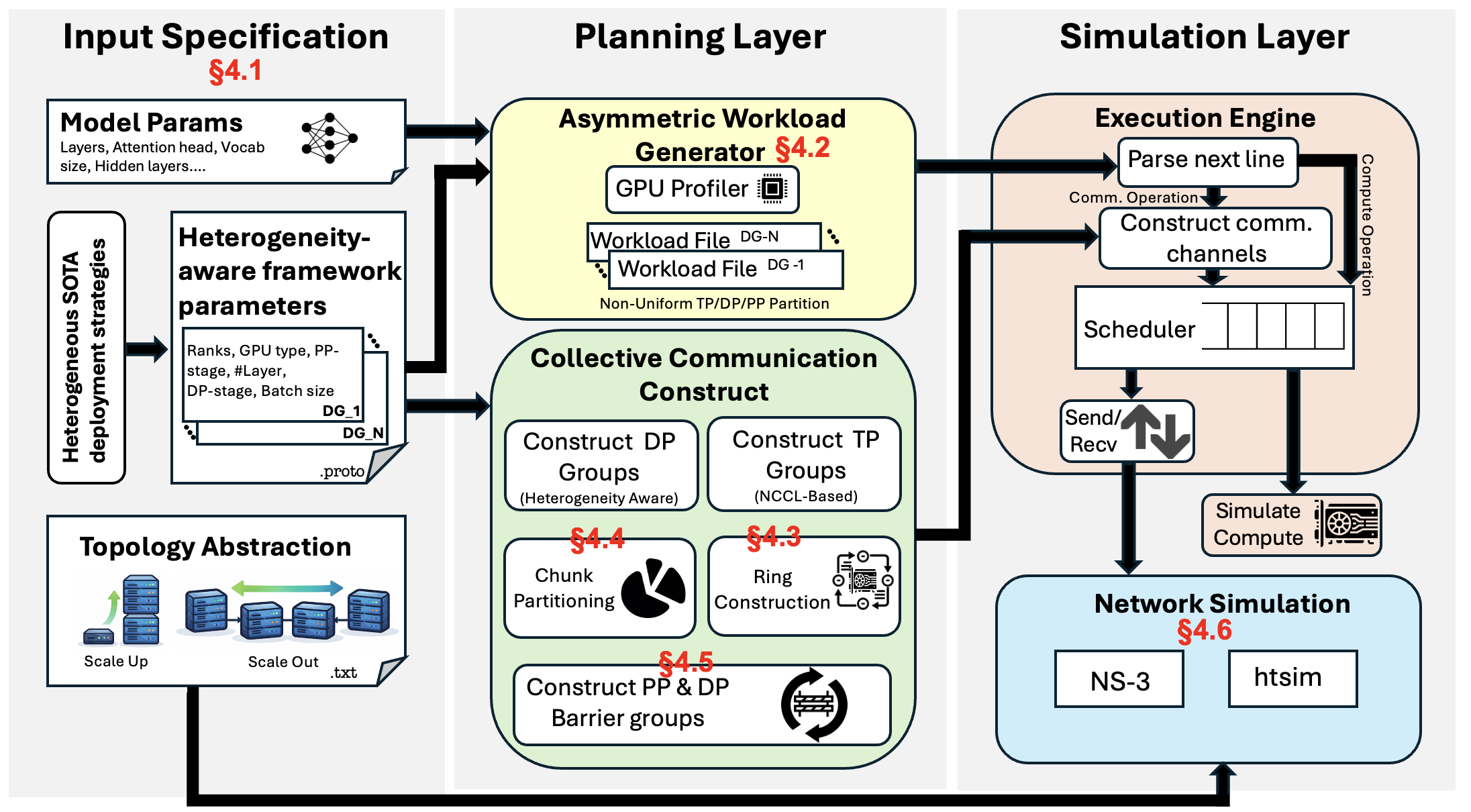}
    \caption{\system's Design }
    \label{fig:design}
\end{figure}

Figure~\ref{fig:design} illustrates~\system's key components distributed across {\em three} layers: (a) {\em Input Specification:} exposes abstractions to ingest heterogeneous cluster configurations generated by heterogeneous training optimization solutions, including multiple device groups (with varying compute, memory), topology (with scale-up and scale-out interconnects), model partitioning, and data partitioning, 
(b) {\em Planning layer:} prepares workload traces that reflect real-world characteristics and constructs heterogeneity-aware communication groups that would be used for gradient synchronization during simulation runtime, and (c) {\em Simulation layer:} uses an event-driven engine to model compute and collective operations (both scale-up and scale-out).

Using the heterogeneity-aware framework parameters, i.e., per-device-group proto files, the planning layer constructs objects (DP/TP groups, PP/DP barrier groups, multi-ring per-DP group) and mappings (model layer to chunk) necessary to simulate collective communication.

The {\em execution engine} parses the workload trace and schedules compute events via a global priority-based event scheduler that supports concurrent per-device execution. For collective operations (e.g., AllReduce), it dynamically instantiates communication channels using the planning-phase configuration and invokes the selected network backend (NS-3 or htsim) through a \texttt{send/recv} interface.

Our~\system implementation spanned $\sim 8300$ lines of code, including NS-3 backend implementation.

\subsection{Input Specification}
\label{subsec:input_spec}
To simulate training in heterogeneous clusters, the user provides as input \textbf{(satisfies [A1] and [A2])}: (1) heterogeneous cluster information, including compute, memory, cluster size, scale-up interconnect, and scale-out interconnect; (2) non-uniform workload to device mapping; and (3) model parameters.
\system consumes deployment plans generated by heterogeneous training planners or specified manually by users. Its input abstraction captures the same logical parameters used by distributed training frameworks such as Megatron-LM~\cite{shoeybi2019megatron} and DeepSpeed~\cite{rasley2020deepspeed}, including TP/DP/PP degrees, communication groups, pipeline stages, layer ranges, micro-batch sizes, and device placement. Heterogeneous deployments (e.g., HexiScale~\cite{yan2024hexiscale}) are represented through a protobuf-based specification that captures non-uniform configurations across device groups, such as GPU type, layer assignment, batch size, and TP degree. We adopt protobuf to align with production ML infrastructure practices (e.g., Ray~\cite{ray_grpc_guide}) and facilitate future integration with deployment pipelines. Network topology, bandwidth, and delay parameters are specified separately through a topology configuration file. Appendix~\S\ref{sec:appendix:input_spec_example} provides example deployment and topology specifications.



\subsection{Asymmetric Workload Generator}
\label{subsec:workload_gen}
Device groups may execute pipeline stages of vastly different depths (e.g., 20 vs. 12 layers), process different micro-batch sizes due to memory constraints, and use different tensor-parallel (TP) degrees. This heterogeneity requires a simulator that can generate and map multiple workloads, rather than broadcasting a single static workload across the cluster. 

\system's Asymmetric Workload Generator (\systemAWG) ingests heterogeneity-aware framework and model parameters to produce distinct workload files for each device group, reflecting each DG's unique computational and communication characteristics.

\systemAWG \textbf{(satisfies [C1])} operates in two phases: 
(a) In the capability profiling phase,~\systemAWG, profiles per-device GPU compute capability (TFLOPS) for each device type in the heterogeneous cluster, using a sample GPU, and
(b) In the trace generation phase,~\systemAWG, extends SimAI's framework interception mechanism. For each device group, the training framework is configured with DG-specific parameters such as assigned layers, micro-batch size, and tensor-parallel (TP) degree, and executed using the profiled compute capabilities. This execution generates a trace of computation and communication operations unique to the device group. These traces precisely capture the operator sequence, tensor shapes, and collective communication patterns executed during training.

\begin{figure}[t]
    \centering
    \includegraphics[width=0.49\textwidth]{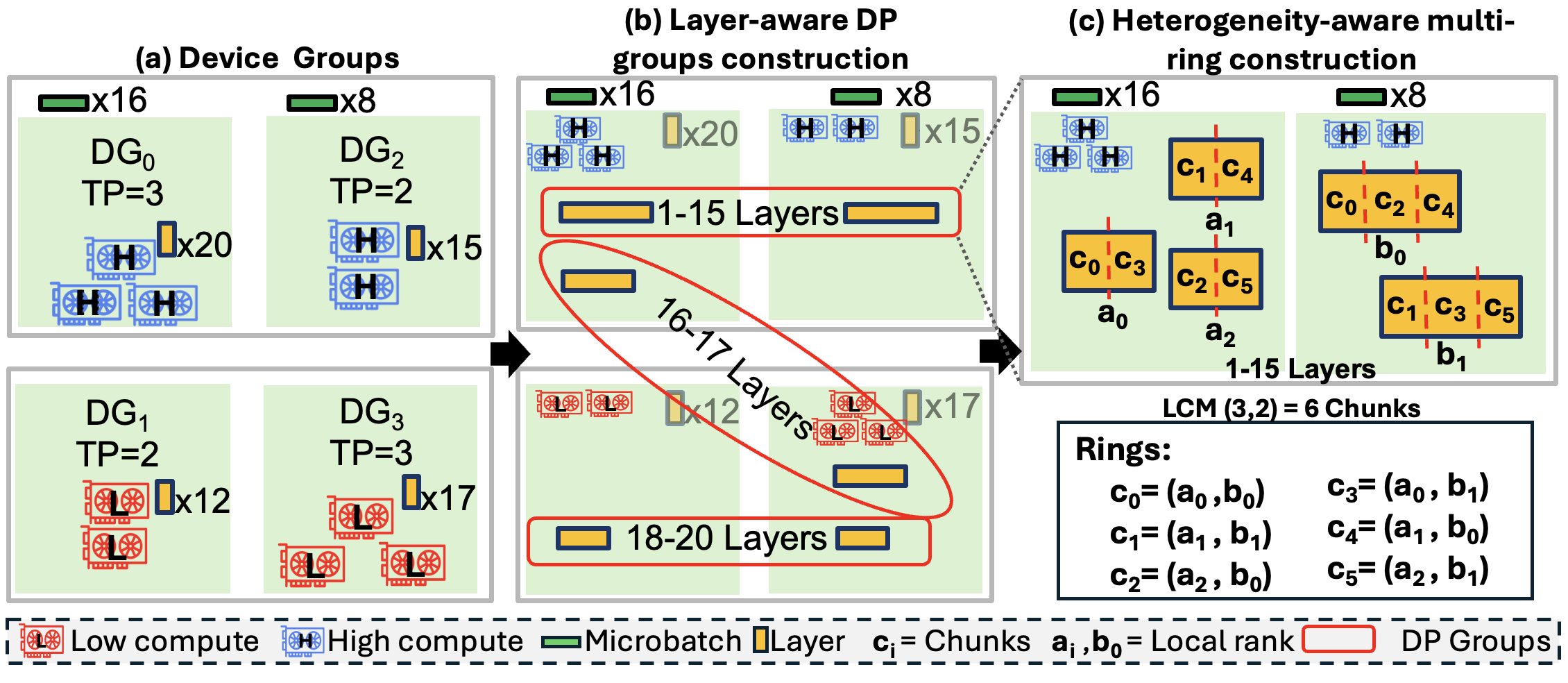}
    \caption{Heterogeneity-aware multi-ring LCM-based resharding in \system for non-uniform layer partitioning across high-compute (blue) and low-compute (red) device groups with layer-aware DP groups. }
    \label{fig:workflow}
\end{figure}

\subsection{Heterogeneity-aware Ring Construction.} \label{subsec:hetero_ring}

Figure~\ref{fig:workflow} shows an example that outlines the workflow for constructing objects and groups required for collective communication. The example shows mixed H100/A100 clusters using non-uniform layer partitioning. High-compute groups (blue) handle more layers (e.g., 20\&15 layers, TP=3\&2), while low-compute groups (red) handle fewer layers, balanced by non-uniform batches.

In heterogeneous clusters, asymmetric pipeline partitioning can cause device groups to process overlapping but non-identical layer ranges. A static DP group would therefore attempt to synchronize gradients for layers not present on all devices, leading to incorrect synchronization. 
In this section, we explain how~\system addresses this challenge.

Using the heterogeneity-aware framework parameters, i.e., per-device-group proto files, DP groups, and a Multi-ring per-DP group is constructed \textbf{(satisfies [C2])}.
First, device groups are used to generate data parallel (DP) groups for gradient synchronization using our {\em Sweep-line algorithm} (see Algorithm~\ref{alg:sweep-line}). We have designed an {\em LCM-based Multi-ring construction} and an {\em LCM-based Gradient chunking} algorithm that generates the multi-ring topology and chunk sizing/mapping for each DP group. 
In Appendix~\S\ref{sec:appendix:multiring_example}, we illustrate the algorithms using a concrete example. 

\begin{algorithm}[t]
\scriptsize
\caption{Sweep-Line Algorithm for Dynamic Group Formation} 
\begin{algorithmic}[1]
    \State \textbf{Input: } Set of Device Groups, $\mathcal{D} = \{DG_1, DG_2, \ldots, DG_n\}$, where each $DG_i$ has assigned layer range, $[s_i, e_i]$
    \State \textbf{Output: } Set of DP synchronization groups, $\mathcal{G}_{DP}$
    \State \textbf{begin:}
    
    \State $\mathcal{P} \leftarrow \emptyset$ 
    \Comment{Stores boundary points for each DP}
    
    \For{each $DG_i \in \mathcal{D}$}
        \State $\mathcal{P} \leftarrow \mathcal{P} \cup \{s_i, e_i + 1\}$ 
    \EndFor
    
    \State $\mathcal{P}_{\text{sorted}} \leftarrow \textsc{Sort}(\mathcal{P})$ 
    \State $\mathcal{P}_{\text{unique}} \leftarrow \textsc{Deduplicate}(\mathcal{P}_{\text{sorted}})$ 
    
    \State $\mathcal{G}_{DP} \leftarrow \emptyset$ 
    \State $\mathit{group\_id} \leftarrow 0$
    
    \For{$i = 0$ to $|\mathcal{P}_{\text{unique}}| - 2$}
        \State $\mathit{seg}_{\text{start}} \leftarrow \mathcal{P}_{\text{unique}}[i]$
        \State $\mathit{seg}_{\text{end}} \leftarrow \mathcal{P}_{\text{unique}}[i+1] - 1$
        
        \State $\mathcal{C} \leftarrow \emptyset$ 
        \Comment{Holds set of DGs that cover $[seg_{start},seg_{end}]$}
        
        \For{each $DG_j \in \mathcal{D}$}
            \If{$s_j \leq \mathit{seg}_{\text{start}}$ and $e_j \geq \mathit{seg}_{\text{end}}$}
                \State $\mathcal{C} \leftarrow \mathcal{C} \cup \{DG_j\}$ 
            \EndIf
        \EndFor
        
        \If{$|\mathcal{C}| \geq 2$} 
            \State $\mathcal{R} \leftarrow \bigcup_{DG_j \in \mathcal{C}} \text{GetRanks}(DG_j)$ 
            \State $G \leftarrow \textsc{CreateDPGroup}(\mathit{group\_id}, [\mathit{seg}_{\text{start}}, \mathit{seg}_{\text{end}}], \mathcal{R})$
            \State $\mathcal{G}_{DP} \leftarrow \mathcal{G}_{DP} \cup \{G\}$
            \State $\mathit{group\_id} \leftarrow \mathit{group\_id} + 1$
        \EndIf
    \EndFor
    
    \State \Return $\mathcal{G}_{DP}$
\end{algorithmic}
\label{alg:sweep-line}
\end{algorithm}

\myparab{Sweep-line algorithm}~\cite{sweepline}.
Consider four Device Groups with the following assigned layer ranges $[s_i,e_i]$:  $DG_0$ with layer range [1,20] and ranks $\{0,1,2\}$, $DG_1$ with layer range [21,32] and ranks $\{5,6\}$, $DG_2$ with layer range [1,15] and ranks $\{3,4\}$, and $DG_3$ with layer range [16,32] and ranks $\{7,8,9\}$. For each Device Group, the sweep-line algorithm first collects all layer boundaries, i.e., $\{1,21,21,33,1,16,16,33\}$ (for algorithmic optimizations, $e_i$ is incremented by one). These boundaries are sorted, and duplicates are removed, resulting in the set $\{1,16,21,33\}$, which defines three unique segments: [1,15], [16,20], and [21,32]. 
For each segment, the algorithm identifies the Device Groups that fully cover the segment and forms a DP group, $\mathcal{G_{DP}}$. 
Segment [1,15] is covered by both $DG_0$ and $DG_2$, so a DP synchronization group is formed that contains the union of their ranks which results in $\{0,1,2,3,4\}$. Similarly, segment [16,20] is covered by $DG_0$ and $DG_3$, forming a second DP group with ranks $\{0,1,2,7,8,9\}$ and lastly, segment [21,32] is covered by $DG_1$ and $DG_3$, which results in a third DP group with ranks $\{5,6,7,8,9\}$.
This decomposition enables Layer-Aware Multi-Group Membership: rank 0 (from $DG_0$) participates in two different DP groups, one for synchronizing layers 1--15 and another for layers 16--20, depending on which layer's gradients are being processed.

The sweep-line algorithm runs once during the planning phase, and it requires $O(D \log D + D \cdot S)$ time, where $D$ is the number of DGs and $S$ is the number of unique layer segments (typically $S \leq 2D$). In realistic deployments, pipeline partitions do not exceed $\approx$ 8 stages~\cite{nvidiadev,narayanan2021efficient}, and many DGs share identical layer boundaries, which keeps $S$ small. As a result, the sweep-line construction scales efficiently in practice, with $S$ bounded by a practical and small constant making the $O(D\cdot S)$ term negligible even at cluster scale.

\begin{algorithm}[t]
\scriptsize
\caption{LCM-Based Multi-Group Construction for Heterogeneous data Synchronization}
\label{alg:lcm-multi-ring}
\begin{algorithmic}[1]
    \State \textbf{Input: } DP synchronization group, $\mathcal{G_{DP}}$, covering layer range $[l_s, l_e]$, with participating device groups, $\{DG_1, DG_2, \ldots, DG_k\}$, having TP degrees, $\{t_1, t_2, \ldots, t_k\}$
    \State \textbf{Output: } $\mathcal{R}$, Multi-ring communication topology for gradient synchronization
    \State \textbf{begin:}
    
    \State $L \leftarrow \text{lcm}(t_1, t_2, \ldots, t_k)$ 
    
    \State $\mathcal{R} \leftarrow \emptyset$ \Comment{\textit{Set of communication rings in a DP group}} 
    
    \For{$c = 0$ to $L - 1$} \Comment{Create one ring per LCM chunk}
        \State $\text{participants}_c \leftarrow \emptyset$ 
        
        \For{each $DG_i \in \{DG_1, DG_2, \ldots, DG_k\}$}
            \State $t_i \leftarrow \text{TP\_degree}(DG_i)$
            \State $\text{ranks}_i \leftarrow \text{GetRanks}(DG_i)$ 
            \For{each rank $r \in \text{ranks}_i$} \Comment{Mapping chunk to a set of local ranks}
                \State $\text{local\_rank} \leftarrow r \mod t_i$ 
                
                \If{($c \mod t_i = \text{local\_rank}$)} 
                    \State $\text{participants}_c \leftarrow \text{participants}_c \cup \{r\}$
                \EndIf
            \EndFor
        \EndFor
        
            \State $\text{Ring}_c \leftarrow \text{CreateCommRing}(c, \text{participants}_c)$
            \State $\mathcal{R} \leftarrow \mathcal{R} \cup \{\text{Ring}_c\}$
    \EndFor
    
    \State \Return $\mathcal{R}$ 
\end{algorithmic}
\end{algorithm}
\myparab{LCM-based Multi-Ring construction algorithm.}
The heterogeneity-aware ring topology is constructed automatically during the planning phase. For each DP synchronization group identified by the Sweep-Line algorithm,~\system's {\em LCM-based Multi-Ring Construction} algorithm (see Algorithm~\ref{alg:lcm-multi-ring}) performs the following steps: 
(1) Generate chunk-to-rank mappings by determining which ranks (across all participating DGs) are responsible for each chunk index, $c \in [0, L-1]$, using the interleaved assignment formula,
(2) instantiate communication rings by creating a MockNccl~\cite{wang2025simai} communicator for each chunk $c$ containing only the ranks assigned to that chunk, configured with the appropriate network topology (e.g., ring, tree, or hierarchical) based on the underlying interconnect structure,
(3) The execution engine stores these mappings in a layer-aware routing table indexed by $(layer\_number, chunk\_index)$, enabling the simulator to quickly look up the correct communication ring during the backward pass.

The LCM-based synchronization mechanism is a critical innovation that enables \system to accurately model the gradient aggregation patterns of state-of-the-art heterogeneous training frameworks such as HexiScale and Metis, which rely on asymmetric TP configurations to balance compute and memory constraints across mixed-generation GPU clusters.

\subsection{Heterogeneity-aware Chunk Partitioning}
\label{subsec:hetero_chunk_post_ring}

With non-uniform workload partitioning, the gradient tensor shapes become incompatible for aggregation. 
\system designs an {LCM-based Gradient Chunking} algorithm 
(see Algorithm~\ref{alg:hetero_chunk_norm})
that repartitions the gradients to establish a common synchronization granularity across the communicating heterogeneous DGs in a DP group. \textbf{(satisfies [C3])}
In Appendix~\S\ref{sec:appendix:chunk_partitioning_example}, we illustrate the algorithms using a concrete example.

The core insight is that if two DGs participating in the same DP group have TP degrees $t_i$ and $t_j$, their gradient chunks can be made compatible by subdividing them into a finer granularity equal to the Least Common Multiple (LCM) of the two TP sizes.
For a DP group spanning layers $[l_s, l_e]$, with TP degrees ${t_1, t_2, \ldots, t_k}$, we set L = $\mathrm{lcm}(t_1, t_2, \ldots, t_k)$.
Given a gradient communication volume, $d$, for the DP group, each device group, $DG_i$, computes its per-rank gradient volume as $\frac{d}{t_i}$, and each rank in $DG_i$ further subdivides the volume into equal-sized chunks, $\frac{L}{t_i}$. Hence, each rank contributes a gradient chunk of size $\frac{d}{L}$ units to each ring it participates in. This guarantees that all communication rings operate on identically sized gradient chunks, ensuring correctness. Algorithm~\ref{alg:hetero_chunk_norm} records these per-chunk, per-rank communication volumes, which are then assigned to ranks using an interleaved strategy, and dedicated communication rings are constructed for each chunk.
We discuss the bounds and tradeoffs of~\system's LCM-based chunking technique in~\S\ref{sec:appendix:lcm_bounds}.

\begin{algorithm}[t]
\scriptsize
\caption{LCM-based Gradient/Data Chunking} 
\label{alg:hetero_chunk_norm}
\begin{algorithmic}[1]
\State \textbf{Input: }DP synchronization group, $\mathcal{G_{DP}}$, with participating device groups, $\{DG_{1}, DG_{2}, \dots, DG_{k}\}$, having TP degrees, $\{t_1,t_2, \dots, t_k\}$, with communication volume, $d$
\State \textbf{Output: }For each $DG_i \in \{DG_1,DG_2, \dots, DG_k\}$ per-chunk per-rank volume
\State $L \gets \mathrm{lcm}(t_1,t_2,\dots,t_k)$ 
\For{each $DG_i \in \{DG_1, DG_2, \ldots, DG_k\}$}
    \State $\text{data\_per\_rank}_{DG_{i}} \gets \dfrac{d}{t_i}$
    \State $\text{chunk\_multiplier}_{DG_{i}} \gets \dfrac{L}{t_i}$
    \State $\text{data\_per\_chunk\_per\_rank} \gets \dfrac{\text{data\_per\_rank}_{DG_{i}}}{\text{chunk\_multiplier}_{DG_{i}}}$ 
\EndFor
\end{algorithmic}
\end{algorithm}

\subsection{Pipeline-Parallel Barrier and Communication Modeling}
\label{subsec:barrier_dp_pp}
\system models stragglers arising from non-uniform workload distribution across pipeline stages \textbf{satisfies ([C4])} and supports hybrid PP--TP--DP execution. To enforce pipeline dependencies, \system creates a {\em Pipeline Barrier Group} (PBG)\footnote{Pipeline Barrier Group (PBG): The set of ranks within a single DP stage.}, ensuring that stages within a model replica execute in order and that later stages wait for preceding stages to complete. Note that the SOTA simulators, such as SimAI~\cite{wang2025simai}, ignore this.  \system further inserts explicit point-to-point activation and gradient transfers between adjacent pipeline stages, with communication groups and chunking derived using the LCM-based synchronization algorithms.

After local computation completes, devices in the {\em DP Barrier Group} (DBG) trigger gradient synchronization across data-parallel replicas. Together, the PBG and DBG abstractions enable faithful simulation of pipeline execution, inter-stage communication, and heterogeneity-induced idle time.

\system provides reusable abstractions for emerging pipeline-parallel algorithms. Beyond the current GPipe implementation, we are extending support to 1F1B~\cite{harlap2018pipedream} and other pipeline schedules.

\subsection{Network Simulation}
\label{subsec:network_simulation}

Accurate performance prediction for heterogeneous LLM training requires modeling the interaction between collective communication and diverse network fabrics. Packet-level simulators such as NS-3 offer high fidelity by modeling per-packet behavior and protocol state, but their cost is prohibitive at scale. A single AllReduce can generate millions of packets, making packet-level simulation of even one training iteration take hours. For example, in our evaluation, simulating one Llama 7B training iteration on 128 GPUs took 2 hours (Figure~\ref{fig:qus1-plot2}). This limits exploration of heterogeneous deployment and network design choices.

\system addresses this scalability gap with dual network backends: NS-3 for protocol-accurate analysis and htsim for fast, scalable simulation. This design lets users trade fidelity for speed, depending on their experimental goals \textbf{(satisfies [C5])}.

\myparab{Extensions to NS-3 for Heterogeneous Clusters.}
NS-3 models the cluster network topology~\cite{bai2024unison}. \system extends NS-3 to capture host-network topology, including scale-up interconnects, PCIe switches, and NIC processing. Packets traverse a detailed path comprising PCIe switches, NICs, Top-of-Rack (ToR) switches, and scale-up and core fabric links, each instantiated with realistic bandwidths and latencies. This model captures mixed-generation deployments (e.g., ConnectX-6 vs. ConnectX-7, PCIe Gen4 vs. Gen5, varying ToR uplink rates), enabling \system to quantify how slower NICs, additional PCIe hops, or reduced bandwidth introduce stragglers and inflate collective completion times. Our NS-3 implementation spanned $\sim 380$ lines of code.

\myparab{Extensions to htsim for Heterogeneous Clusters.}
\system integrates {\em htsim}~\cite{htsim} as a network backend that models communication at a flow-level abstraction for speedup. 
We design and implement {\em three} extensions to htsim to enable simulation of heterogeneous host network topology.

\begin{figure}[t]
    \centering
    \includegraphics[width=0.4\textwidth]{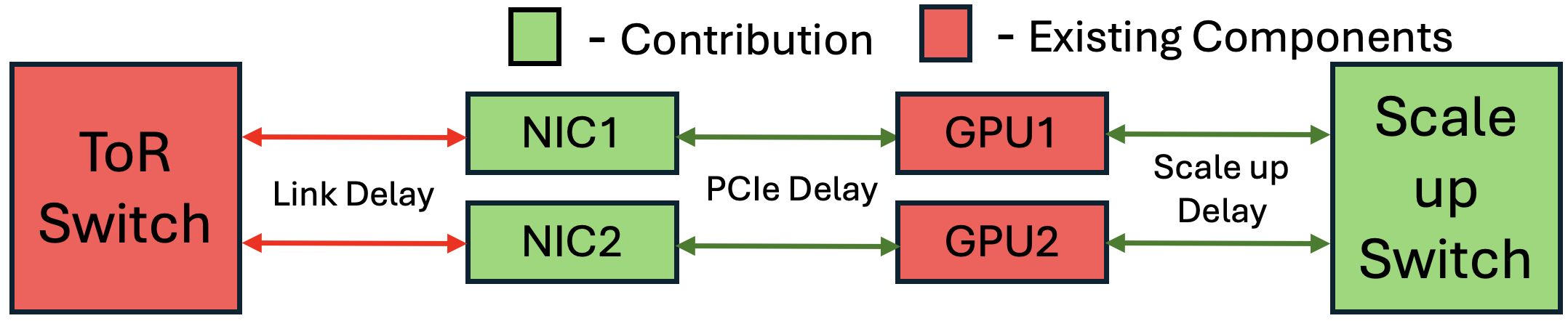}
    \caption{Overview of contributions to htsim implementation.}
    \label{fig:htsim_impl}
\end{figure}

\noindent{\em \textbf{(1) PCIe switch model.}}
In the htsim~\cite{htsim} design, GPUs are directly connected to the ToR switches. To simulate PCIe switching cost, we implemented an additional layer of switches (using htsim's {\em FatTreeSwitch} class) between the GPUs and ToR, which acts as the {\em NIC} (see Figure~\ref{fig:htsim_impl}). GPU--NIC and NIC--ToR are implemented as one-to-one and one-to-many connections, respectively. GPU--NIC interconnect is configured with real-world bandwidth and processing delays to simulate PCIe costs for heterogeneous nodes, while the remainder of the scale-out topology remains unchanged.

\noindent {\em \textbf{(2) Scale-up network model.}}
To model intra-node, inter-GPU communication, we designed and implemented a dedicated scale-up network layer (see Figure~\ref{fig:htsim_impl}) (using {\em FatTreeSwitch} htsim class).
To represent the scale-up fabric, the GPUs connect to a scale-up switch configured with low latency (sub-µs) and high bandwidth (say, 600–900 GB/s). 
We implemented the htsim function, {\em get\_bidir\_paths()} that updates routing tables such that GPUs within the same machine connect to the same scale-up switch, bypassing ToR and Aggregator (AGG) switches for TP and intra-node PP communication. The scale-up switch models scale-up non-blocking behavior with negligible queueing delay.

\noindent {\em \textbf{(3) Rail-Optimized topology integration for training clusters.}}
The current htsim implementation does not support the rail-optimized topology commonly used in AI training clusters. {\em htsim} models a fat-tree topology, which restricts the network to ToR and aggregation switches with full all-to-all connectivity. To isolate bandwidth across collective traffic, a rail-optimized topology connects GPUs with identical local ranks to the same ToR via dedicated NICs, forming isolated rails that bypass aggregation switches during collective operations such as AllReduce.

We extend {\em htsim} to model rail-optimized topology by partitioning the network graph into isolated rail subgraphs. To model this scale-out network, we implemented the htsim function, {\em get\_bidir\_paths()}, which updates the routing table for both the scale-out and scale-up networks. 
This design scales to large dataplane elements while accurately capturing rail isolation, where GPUs of the same local rank share dedicated ToR channels. Our {\em htsim} implementation spanned $\sim 500$ lines of code.

Our htsim prototype can independently be integrated with trace-driven simulators such as ATLAHS~\cite{shen2025atlahs} that support generic workloads, i.e., AI, as well as HPC

\section{Evaluation} 
\label{sec:eval}

We run Llama 2 (7B, 13B) and GPT-175B models across experiments, with fixed optimizers and controlled DP/TP/PP settings to isolate each evaluation dimension. 

\myparab{Simulation testbed.} 
We run~\system and SimAI simulation frameworks on {\em two}
32-core AMD EPYC-9354 nodes with $2 \times A100$ and $4 \times H100$ GPUs. We generate model workloads using the AICB workload benchmark~\cite{aicb_benchmark} on A100 and H100 GPUs to run simulations for homogeneous and heterogeneous configurations.
All experiments were run for one training iteration, unless mentioned otherwise. Table~\ref{tab:gpu_specific_delay_table} shows the heterogeneous cluster's compute and interconnect specifications, which we used for simulation-only experiments.  

\myparab{Real-world testbed.} 
We evaluate \system on a heterogeneous testbed comprising: (i) two AMD EPYC 9354 nodes, each with $2\times$A100 and $4\times$H100 GPUs, scale-up communication uses PCIe Gen4/Gen5 interconnects, while scale-out communication uses Intel X710 and X550T NICs. (ii) an Intel Xeon 8480C node with $8\times$H200 GPUs, and an Intel Xeon 8570 node with $8\times$B200 GPUs, with NVLink for scale-up and ConnectX-7 for scale-out, as summarised in Table~\ref{tab:hetero-config}. The nodes are distributed across racks connected through ToR and aggregation switches. Results are averaged over three runs of a single training iteration unless stated otherwise. Unless specified, training-time prediction experiments use the NS-3 backend.


\begin{table}[t]
\centering
\tiny
\caption{Cluster Configuration. Column~2 shows the cluster size and GPU layout (e.g., $2 \times (4 \times \text{H100})$ means two nodes with four H100 GPUs each). Column~3 shows model parallelism strategy (e.g., $4 \times ((4 \times \text{TP})\text{-DP})$ means four data-parallel replicas, each using 4-way tensor parallelism)}
\label{tab:config-model-cluster}
\renewcommand{\arraystretch}{1.2}
\setlength{\tabcolsep}{4pt}
\begin{tabular}{|c|c|c|}
\hline
\textbf{Config}  & \textbf{Compute Cluster}  & \textbf{Model Distribution} \\
\hline
C1 & 2 $\times$ H100    & DP \\
\hline
C2 & 2 $\times$ A100   & DP \\
\hline
C3 & 2 $\times$ (4 $\times$ H100) & DP \\
\hline
C4 & 2 $\times$ (4 $\times$ A100) & DP \\
\hline
C5 & 2 $\times$ (4 $\times$ H100) & 2 $\times$ ((4 $\times$ TP) - DP) \\
\hline
C6 & 2 $\times$ (4 $\times$ A100) & 2 $\times$ ((4 $\times$ TP) - DP) \\
\hline
C7 & 4 $\times$ (4 $\times$ H100) & 4 $\times$ ((4 $\times$ TP) - DP) \\
\hline
C8 & 4 $\times$ (4 $\times$ A100) & 4 $\times$ ((4 $\times$ TP) - DP) \\
\hline
C9 & 1 $\times$ A100, 1 $\times$ H100  & DP \\
\hline
C10 & 2 $\times$ A100, 2 $\times$ H100  & A (2 $\times$ DP) - DP - H (2 $\times$ DP) \\
\hline
C11 & 2 $\times$ A100,  2 $\times$ H100  & A (2 $\times$ TP) - DP - H (2 $\times$ TP) \\
\hline
C12 & 2 $\times$ A100,  2 $\times$ H100  & A (2 : 2 PP) - DP - H (2 : 2 PP) \\
\hline
C13 & 4 $\times$ A100, 4 $\times$ H100 & A (4 $\times$ DP) - DP - H (4 $\times$ DP) \\
\hline
C14 & 4 $\times$ A100, 4 $\times$ H100 & A (4 $\times$ TP) - DP - H (4 $\times$ TP) \\
\hline
C15 & 4 $\times$ A100, 4 $\times$ H100 & A (3 $\times$ TP : 1) PP - DP - H (3 $\times$ TP : 1) PP \\
\hline
C16 & 2 $\times$ (4 $\times$ A100), 2 $\times$ (4 $\times$ H100) & 2 $\times$ (A (4 $\times$ TP) - DP - H (4 $\times$ TP)) \\
\hline
\end{tabular}
\end{table}

\begin{table}[t]
\centering
\tiny
\caption{Real-world Cluster Interconnect Configuration.} 
\label{tab:hetero-config}
\renewcommand{\arraystretch}{1.2}
\setlength{\tabcolsep}{3pt}
\begin{tabular}{|c|c|c|c|c|c|c|}
\hline
\textbf{GPU} &
\makecell{\textbf{Scale-Up}\\\textbf{BW}} & \makecell{\textbf{Scale-Up}\\\textbf{Delay}} & \makecell{\textbf{PCIe}\\\textbf{BW}} & \makecell{\textbf{PCIe}\\\textbf{Delay}} & \textbf{NIC BW} & \textbf{NIC delay} \\
\hline
A100 40GB~\cite{nvidia_a100} & 2400 Gb/s & 30.66 ns & 256 Gb/s & 2×287.5 ns & 10 Gb/s & 3 µs \\
\hline
H100 80GB~\cite{nvidia_h100} & 3600 Gb/s & 20.44 ns & 512 Gb/s & 2x143.75 ns &10 Gb/s & 1.5µs \\
\hline
H200 141GB~\cite{nvidia_h200_2026} & 3600 Gb/s & 20.44 ns & 512 Gb/s & 2x143.75 ns & 200 Gb/s & 368 ns \\
\hline
B200 192GB~\cite{nvidia_dgx_b200_2026} & 7200 Gb/s & 10.22 ns & 512 Gb/s & 2x143.75 ns & 200 Gb/s & 368 ns \\
\hline
\end{tabular}
\end{table}

\begin{table*}[t]
\centering
\caption{Cluster configurations used for simulation-only experiments.}
\label{tab:gpu_specific_delay_table}
\scriptsize
\resizebox{\textwidth}{!}{%
\begin{tabular}{|c|c|c|c|c|c|c|c|}
\hline
\textbf{Architecture} & \textbf{GPU}  & \textbf{\makecell{NVLink BW\\(Gbps)}} & \textbf{\makecell{NVLink delay\\(ns)}} & \textbf{\makecell{PCIe BW\\(Gbps)}} & \textbf{\makecell{PCIe latency\\(ns)}} & \textbf{\makecell{NIC BW\\(Gbps)}} & \textbf{\makecell{NIC processing delay\\(ns)}} \\
\hline
\makecell{Ampere~\cite{connectx7,nvidia_a100}} &\makecell{ A100 (40GB)} & \makecell{2400 (Gen 3)} & 30.66 & \makecell{256 (Gen 4)} & \makecell{$2\times287.5$} & \makecell{400 (ConnectX-7)} & 368 \\
\hline
\makecell{Hopper~\cite{connectx7,nvidia_h100} }&\makecell{ H100 (80GB)} & \makecell{3600 (Gen 4)} & 20.44 & \makecell{512 (Gen 5)} & \makecell{$2\times143.75$} & \makecell{400 (ConnectX-7) } & 368 \\
\hline
\end{tabular}
}
\end{table*}

\begin{figure*}[t]
\centering
\captionsetup{font=small}
\begin{minipage}[t]{0.30\textwidth}
    \centering

    \includegraphics[width=\linewidth]{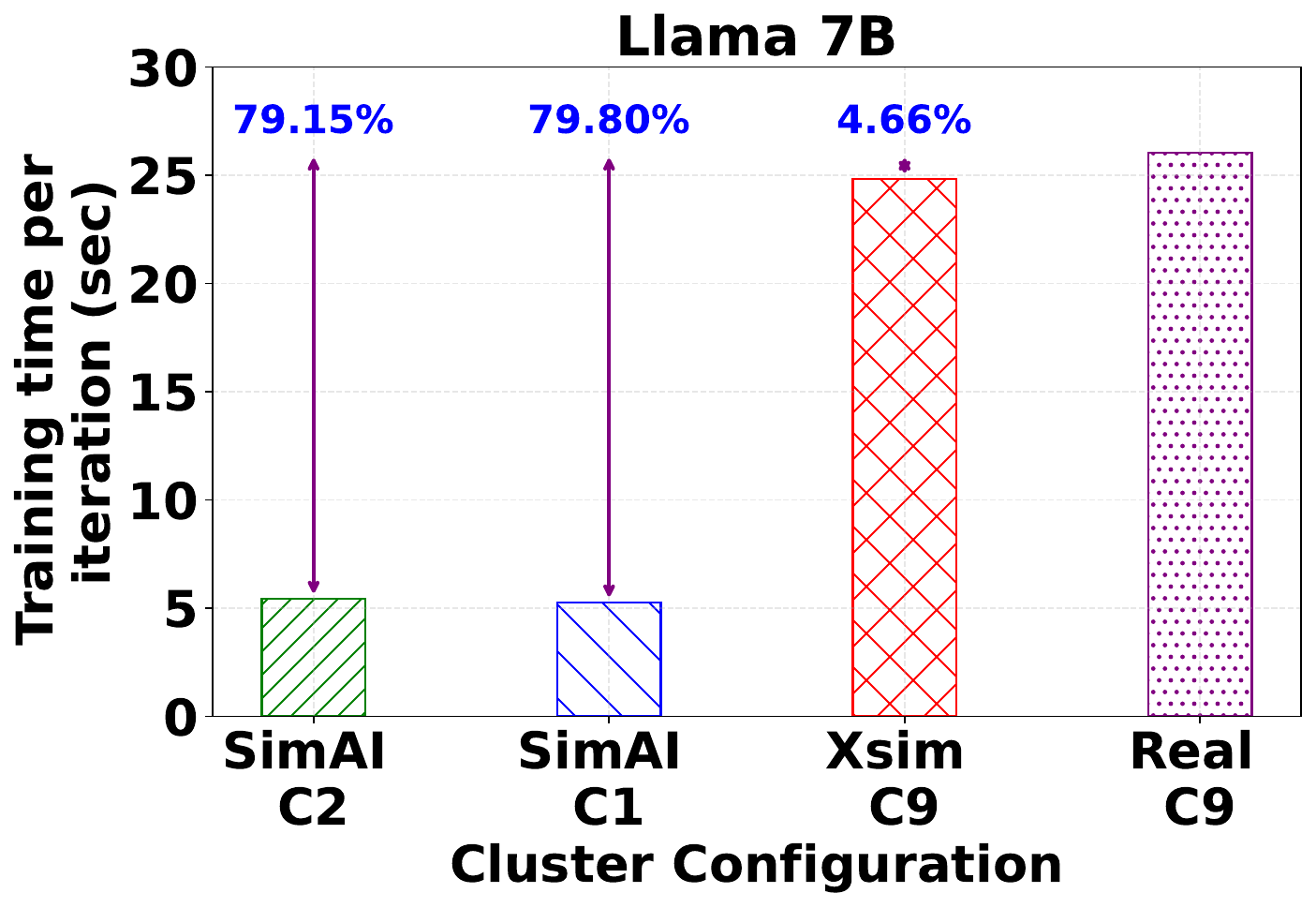}
    \caption{Training time per iteration for Llama 7B on a heterogeneous cluster shows that \system closely matches real hardware with <5\% error, while SimAI incurs large errors due to explicit heterogeneity modeling.
    }
   
    \label{fig:qus2-plot}

\end{minipage}
\hfill
\begin{minipage}[t]{0.32\textwidth}
    \centering
    \captionsetup{font=small}
    \includegraphics[width=\linewidth]{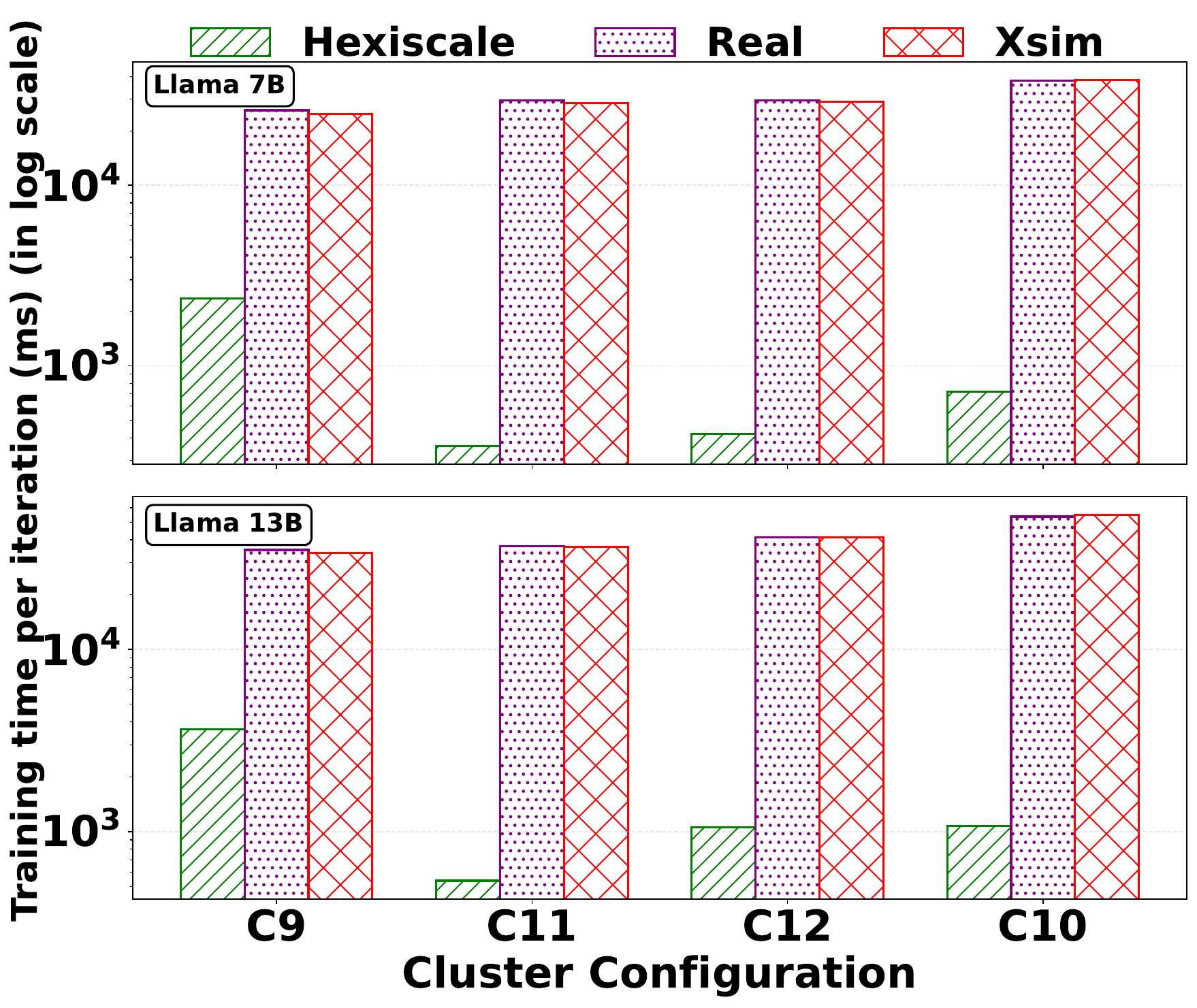}
    \caption{Training time per iteration for Llama 7B and 13B across heterogeneous cluster configurations, comparing Hexiscale, \system, and real hardware. 
    }
    \label{fig:qus3-plot}
\end{minipage}
\hfill
\begin{minipage}[t]{0.30\textwidth}
    \centering
    \includegraphics[width=\linewidth]{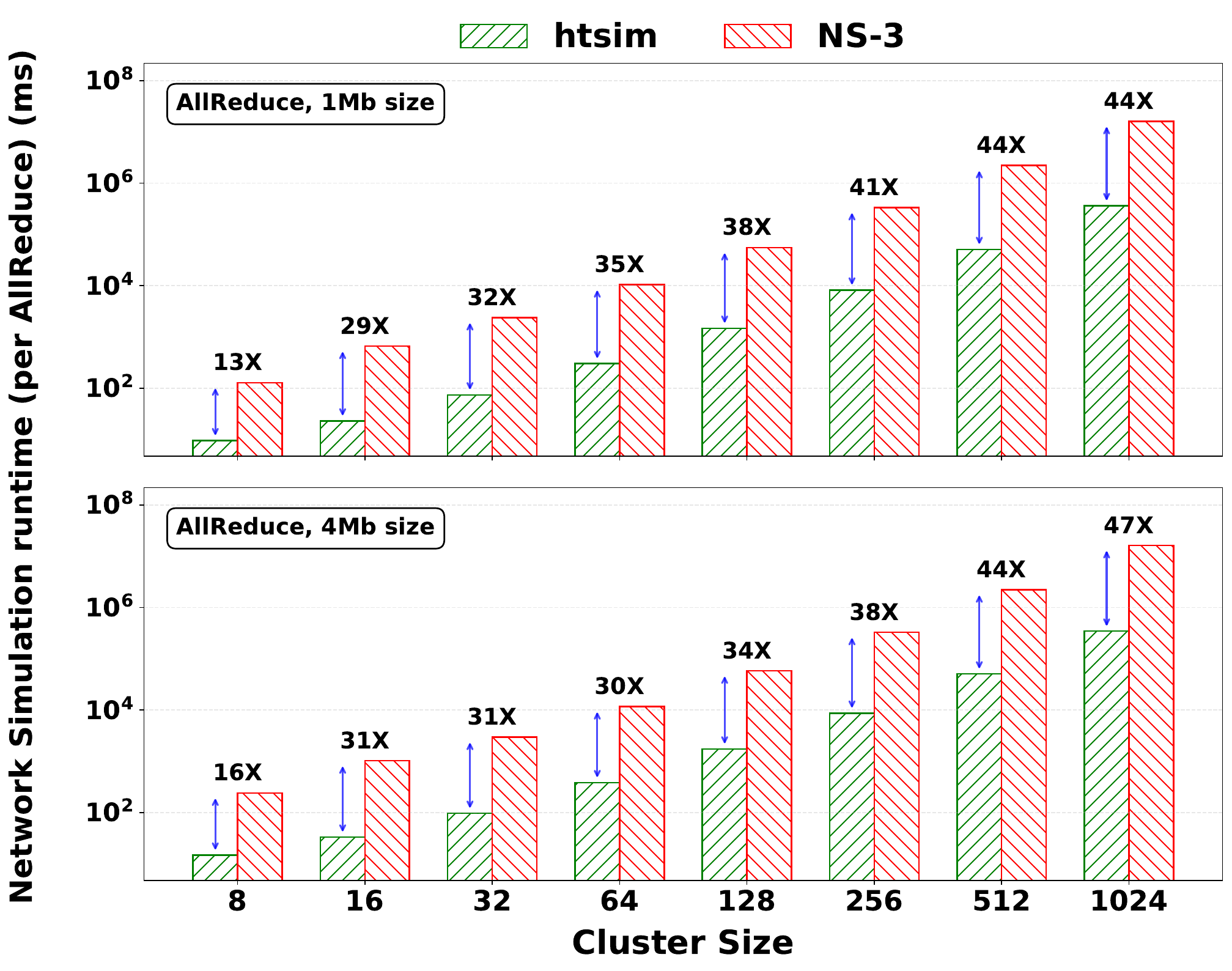}
    \caption{Cluster size vs. Network simulation runtime. NS-3's protocol execution incurs higher overhead at larger scales and smaller message sizes ($16\times$–$47\times$).}
    \label{fig:qus5-plot}
\end{minipage}
\end{figure*}

\myparab{Metrics.}
We measure (1) {\em Prediction accuracy} as the percentage error between the simulated training time (\system, SimAI~\cite{wang2025simai}, or Hexiscale~\cite{yan2024hexiscale}) and real hardware execution time, percentage error of scale-up (TP communication) and scale-out (DP communication using multi-ring) implementation with a real cluster (2) {\em Simulation runtime} as the wall-clock time required to complete a full training iteration simulation, (3) {\em GPU idle time} (a.k.a., Straggler waiting time) measures the GPU waiting time due to {\em AllReduce} completion skew across DP groups in a heterogeneous cluster, and (4) {\em Total Cost of Ownership (TCO)} represents capital expenditure on GPU hardware, and it is computed as $\text{TCO} =\text{CapEx}/TrainingTime$, where {\em CapEx} is the GPU cost in USD, with the unit, ``\$/GPU-hour", (5) Pipeline bubble time is the GPU waiting time during pipeline resharding and communication.

\myparab{Evaluation Questions.}

\noindent{(1) } Can we use existing training simulators that assume homogeneous AI clusters to simulate heterogeneous workload? 
    
\noindent{(2) } How accurately does~\system predict real-world training performance on heterogeneous clusters compared to existing heterogeneity-aware simulators?  


\noindent{(3) } How does the scalability of htsim versus NS-3 affect simulation time in~\system? 


\noindent{(4)} How accurately does~\system predict collective communication overheads across scale-up fabric, i.e., NVLink?

\noindent{(5)} How accurately does~\system predict the time taken for Data Parallelism (DP) multi-ring collective communication operations across different model scales on a heterogeneous distributed cluster?

\noindent{(6)} What is the source of prediction error in~\system?

\noindent{(7)} How does~\system perform compared to the SOTA resharding solutions, Hetauto and AlpaComm, with varying topology abstractions?

\noindent{(8) } How does~\system compare with state-of-the-art training simulators for homogeneous AI clusters? (with respect to prediction accuracy and simulation runtime) (\S\ref{sec:evaluation-extended})

\noindent{(9) } How well does~\system scale with model and cluster size? 
 (\S\ref{sec:evaluation-extended})

\noindent{(10) } How effectively does~\system support system and capacity planning by exposing heterogeneity-aware metrics such as {\em GPU idle time} (due to stragglers) and Total Cost of Ownership (TCO) across cluster designs?(\S\ref{sec:evaluation-extended})




    
    


\myparab{(1) Comparing heterogeneous deployment's training time in the wild (i.e., the ground truth), with SimAI and~\system.}
We run the heterogeneous deployment, $C9$ (see Table~\ref{tab:config-model-cluster} and~\fref{fig:qus2-plot}), i.e., 1 $\times$ A100, 1 $\times$ H100, over the real-world cluster, and also simulate it using~\system. 
Since SimAI does not support heterogeneous system architecture, we deploy $C9$ over two homogeneous configurations, $C1$ (2 $\times$ H100) and $C2$ (2 $\times$ A100). 
Results show that SimAI incurs large prediction errors (~80\%).
This is because SimAI assumes homogeneous AI clusters and cannot accurately simulate collective communication for non-uniform model parallelism, device bandwidth, delays, and communication asymmetry.
\system, by explicitly modeling heterogeneity and workload transformation, achieves near-real accuracy (<5\% error). 

\begin{figure*}[t]
\centering
\captionsetup{font=small}
\begin{minipage}[t]{0.32\textwidth}
    \centering
    \includegraphics[width=\linewidth]{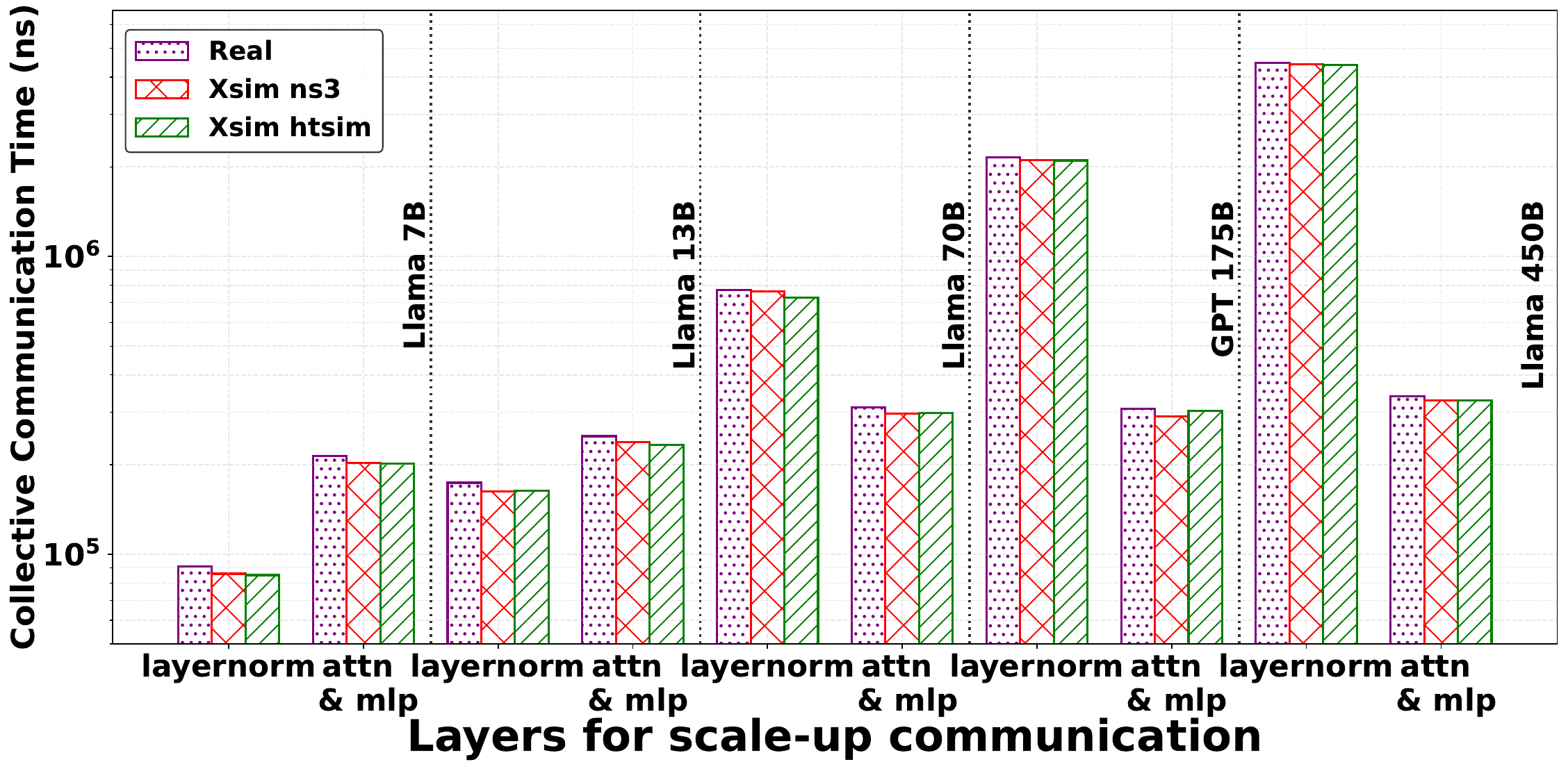}
    \caption{ Comparison of isolated scale-up TP collective communication time between \system (using NS-3 and htsim backends) and real cluster of 8$\times$ H200 NVLink nodes with an average error of 5.5\%. }
    \label{fig:qus7_plot}

\end{minipage}
\hfill
\begin{minipage}[t]{0.32\textwidth}
    \centering
    \includegraphics[width=\linewidth]{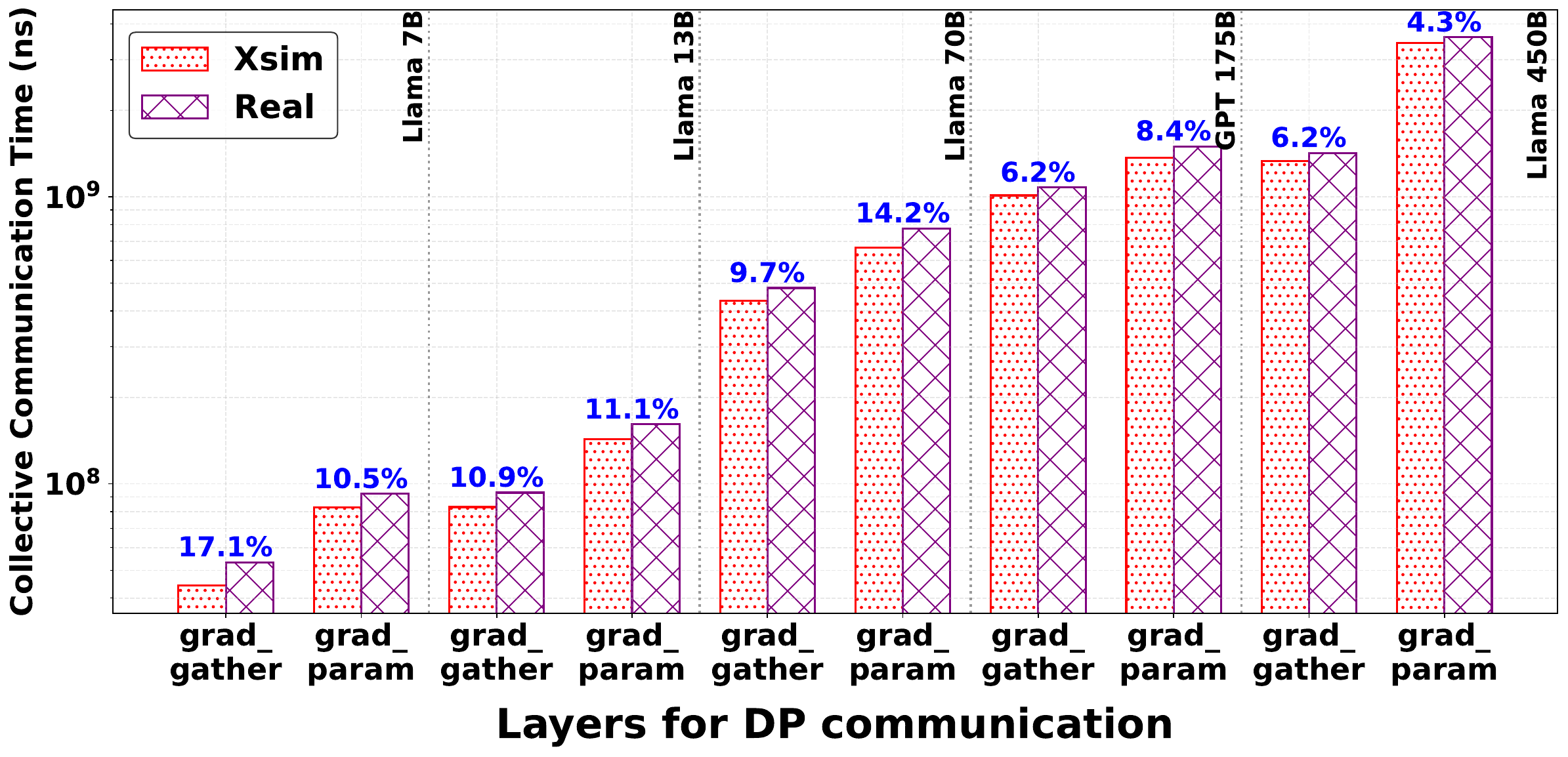}
    \caption{ Comparison of scale-out DP collective communication time between \system and a real cluster making the multi-ring on a non-isolated cluster of 4$\times$ H100 \& 2$\times$ A100 cluster. }
    \label{fig:qus8_plot}

\end{minipage}
\hfill
\begin{minipage}[t]{0.35\textwidth}
    \centering
    \includegraphics[width=\linewidth]{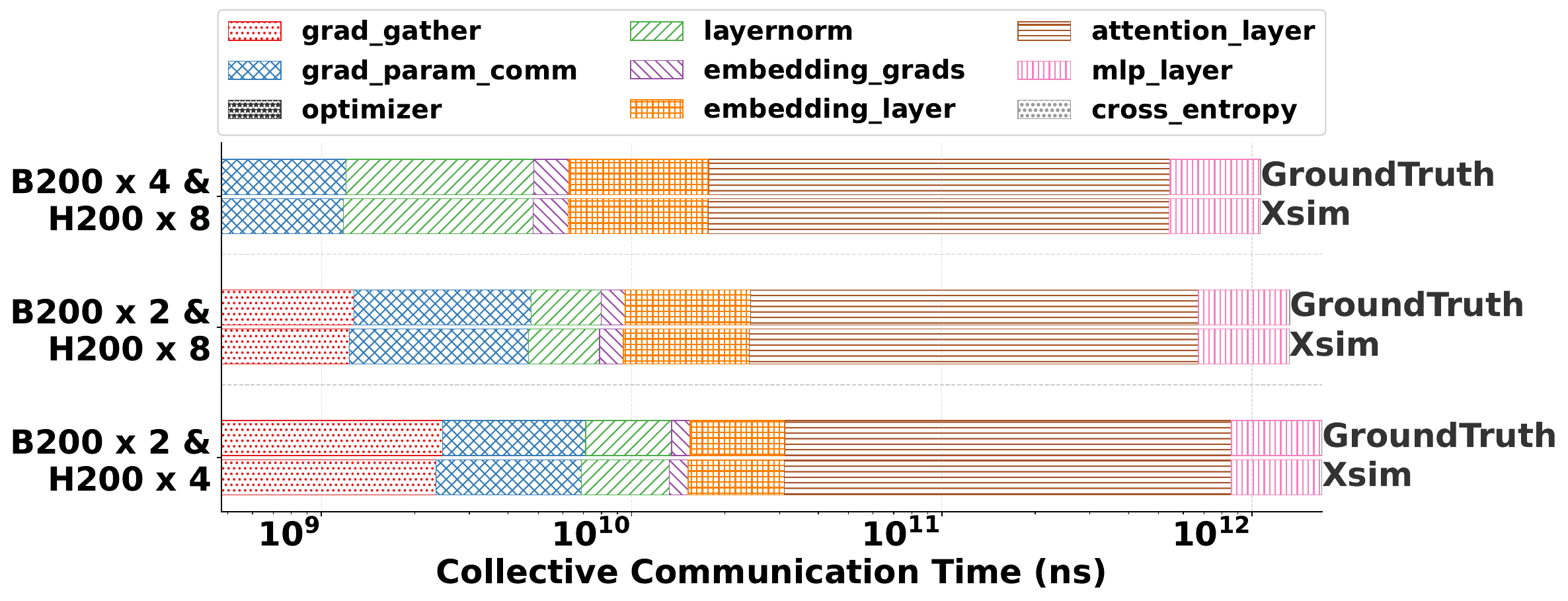}
    \caption{ Layer-wise validation of execution time across diverse heterogeneous cluster configurations. The breakdown isolates of layers, demonstrating \system's precision down to individual layers}
    \label{fig:qus9_plot}

\end{minipage}
\end{figure*}

\myparab{(2) Comparing heterogeneous deployment's training time in the wild (i.e., the ground truth), with Hexiscale ' s~\cite {yan2024hexiscale} heterogeneity-aware analytical simulator, and~\system.}
\fref{fig:qus3-plot} demonstrates the training time for Hexiscale,~\system, and real-world evaluations, across {\em four} heterogeneous configurations, $C9$, $C10$, $C11$, and $C12$ (see Table~\ref{tab:config-model-cluster}). These configurations comprise heterogeneous compute, interconnect, and non-uniform hybrid parallelism strategies. 
For both Llama 7B and Llama 13B, we observe that Hexiscale underestimates training time by 90–99\%, since its analytical model does not account for the synchronization overheads introduced by asymmetric communication during collective operations. In contrast,~\system consistently achieves near-real accuracy, with <5\% error in most configurations, demonstrating its effectiveness as a full-stack heterogeneity-aware training simulator. 
This experiment highlights the fact that network communication cannot be ignored in large-scale AI training clusters.

\myparab{(3) Network simulation scalability: NS-3 vs. {\em htsim}.}
\fref{fig:qus5-plot} shows how~\system's network simulator backends, NS-3~\cite{bai2024unison} (uses $8$ CPU cores) and htsim (uses $1$ CPU), scale with increasing cluster size. We evaluate the simulators over the basic communication unit in the AI training cluster, i.e., {\em AllReduce}, with {\em two} sizes, $1Mb$ and $4Mb$. 
NS-3 provides packet-level fidelity with protocol execution, hence its simulation overhead increases with cluster size, particularly under small chunk sizes and large data volumes. {\em htsim} is $16\times$--$44\times$ faster than NS-3 in the cluster from 8 nodes to 1024 nodes, since htsim uses flow-level abstraction and abstracts away the packet header parsing.

A simulator user interested in protocol-level analysis must use NS-3. However, network traffic analysis, such as congestion control, and microarchitectural analysis, such as scale-up switch queueing delays, can be analyzed using htsim.

\myparab{(4) How accurately does~\system predict scale-up collective communication overhead? } 
To evaluate the predictive fidelity of \system's scale-up network models, we measure TP collective communication time on an 8$\times$H200 NVLink node across models ranging from Llama-7B to Llama-450B and GPT-175B. Figure~\ref{fig:qus7_plot} compares the NS-3 and htsim backends against real hardware measurements for LayerNorm and Attention/MLP collectives. For the large-message Attention/MLP collectives, both backends closely track hardware performance, with average error below 5.5\%. The small remaining error is primarily attributable to NCCL runtime overheads and htsim's flow-level abstraction. These results validate \system's modeling of scale-up communication while demonstrating that htsim provides a scalable alternative when packet-level fidelity is not required.

\myparab{(5) How accurately does \system predict the time taken for Data Parallelism (DP) multi-ring collective communication operations across different model scales on a heterogeneous distributed cluster?} 
To evaluate \system's multi-ring chunk-partitioning algorithm in a scale-out setting, we measure DP synchronization time on a heterogeneous cluster comprising $4\times$H100 and $2\times$A100 GPUs across models ranging from Llama-7B to Llama-450B and GPT-175B. Figure~\ref{fig:qus8_plot} compares simulated and real execution times for gradient-gather and gradient-parameter collectives. \system closely tracks hardware performance across all models, with error decreasing as model size increases. The remaining discrepancy at large scales stems from NCCL implementation constraints: while \system abstracts multi-ring communication as fully parallel chunk transfers, real NCCL executions are limited by link assignment and resource contention, which can serialize communication across rings~\cite{nvidia_nccl_2018}.

\myparab{(6) What is~\system's source of error in the prediction of execution and communication time compared to real-world execution times? -- a layer-wise Ablation study.} 
To evaluate \system's fine-grained fidelity, we compare layer-wise execution times across three heterogeneous clusters: $B200\times4$ \& $H200\times8$, $B200\times2$ \& $H200\times4$, and $B200\times2$ \& $H200\times8$. Figure~\ref{fig:qus9_plot} shows that ~system closely matches measured execution times across compute, communication, and optimizer operations spanning three orders of magnitude. Compute-intensive layers such as \texttt{attention\_layer} and \texttt{mlp\_layer} exhibit less than 1\% error, while communication operations (\texttt{grad\_gather} and \texttt{grad\_param\_comm}) remain within 
1--4\% error across all hardware configurations. These results demonstrate that~\system accurately captures both computation and heterogeneous scale-out communication behavior at layer granularity.

\begin{figure}[t]
    \centering
    \includegraphics[width=\linewidth]{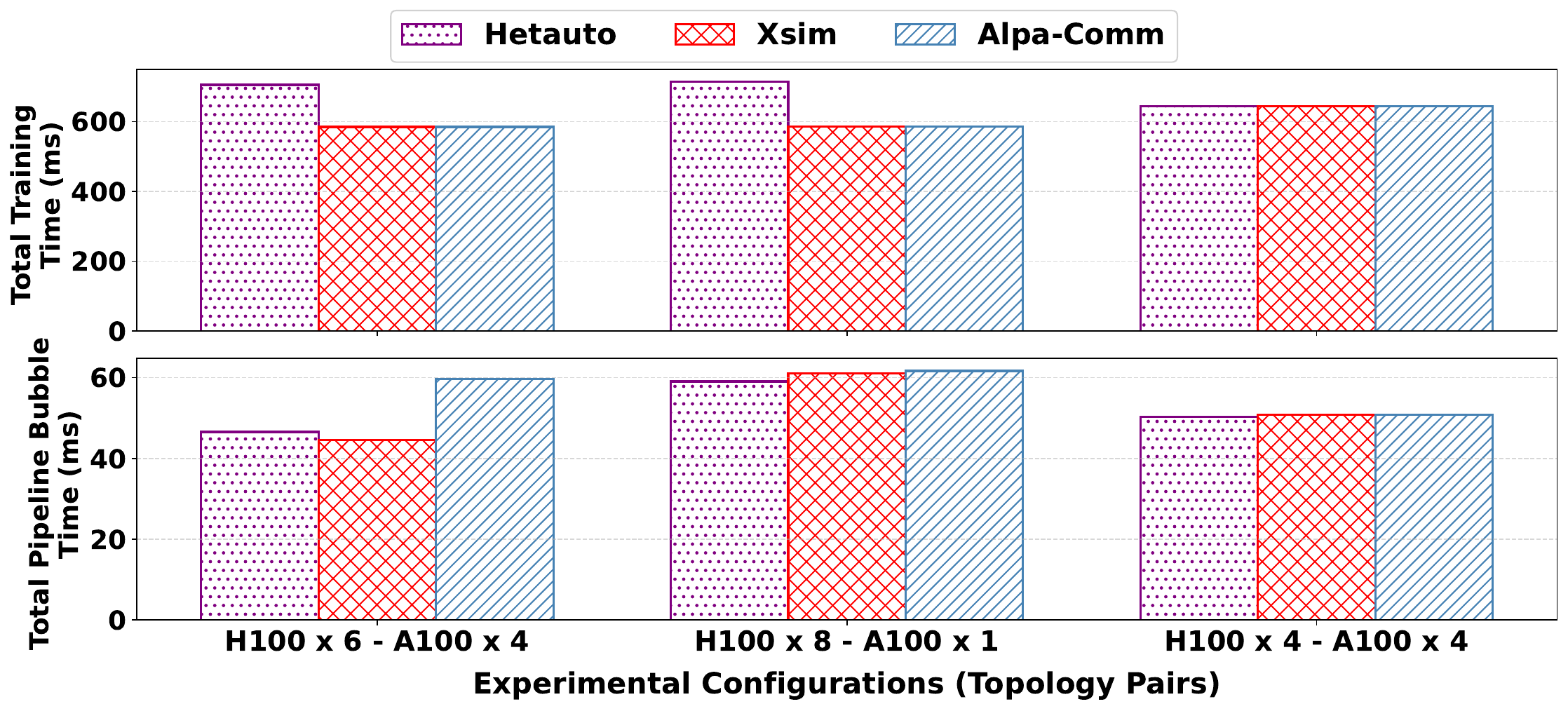}
    \caption{Total training time \& exposed PP communication time comparison across asymmetric topology pairs comparing different re-sharding technique algorithms.}
    \label{fig:qus10_plot}
\end{figure}

\myparab{(7) How does \system perform compared to the SOTA resharding solutions with varying topology abstractions?} 
We compare~\system's predicted performance with two state-of-the-art resharding schemes, AlpaComm~\cite{zhuang2023optimizing} and HetAuto~\cite{qi2026hetauto}, across heterogeneous deployments (recall~\S\ref{subsec:tensor_resharding_sota}): $H100\times6 \rightarrow A100\times4$, $H100\times8 \rightarrow A100\times1$, and $H100\times4 \rightarrow A100\times4$. Figure~\ref{fig:qus10_plot} reports total training time and exposed pipeline bubble time. HetAuto incurs higher training time in asymmetric configurations due to its hierarchical gather–P2P–scatter workflow, whose benefits diminish as the GCD between source and destination parallelism degrees decreases. AlpaComm and~\system avoid these additional aggregation phases through point-to-point communication, reducing end-to-end training time. For symmetric configurations, all three schemes achieve similar performance.

For pipeline bubble time,~\system and HetAuto partition tensors into balanced communication units, resulting in lower stage imbalance and idle time. In contrast, AlpaComm's cutpoint-union approach produces non-uniform communication slices, leading to load imbalance, communication skew, and higher pipeline bubble time.

\section{Related work}

\myparab{Optimizations in the presence of heterogeneous distributed training infrastructure.}
Prior heterogeneous AI training systems~\cite{park2020hetpipe,ye2024asteroid,um2024metis,jia2022whale,zhang2023pipepar,wu2025hetermoe,yan2024hexiscale,zhang2024hap} optimize hybrid parallelism and workload placement across heterogeneous compute and network resources. Learning-based~\cite{yi2020optimizing,liu2023heterps} and analytical planners~\cite{strati2025sailor,mei2025helix} further automate deployment decisions, but abstract away detailed communication and network dynamics. \system complements these works by enabling full-stack simulation of heterogeneous compute, communication, and network configurations.


\myparab{Heterogeneity-aware simulators.}
Heterogeneity-aware simulators such as LLMServingSim~\cite{cho2024llmservingsim} and HeteroSim~\cite{yue2026heterosim} focus on inference or assume uniform workload partitioning, limiting support for heterogeneous training. Similarly, Wu et al.~\cite{wu2025rethinking} use SimAI~\cite{wang2025simai} for heterogeneous planning without extending it to model heterogeneous workload assignment or collective communication. In contrast,~\system supports non-uniform workload partitioning, heterogeneous synchronization, and full-stack training simulation.

\myparab{Heterogeneity-aware collective communication}
To address cluster topology and link bandwidth heterogeneity, HeteCCL~\cite{heiheteccl} and ForestColl~\cite{zhao2024forestcoll} design efficient scheduling solutions. 
\system acts as a bridge between heterogeneous compute deployment works, such as Metis~\cite{um2024metis}, and these works.

\myparab{Resharding Across Heterogeneous Tensor Layouts.}
HetAuto~\cite{qi2026hetauto} addresses tensor-layout mismatches using a GCD-based {\em gather–P2P–scatter} resharding strategy across heterogeneous pipeline stages. AlpaComm~\cite{zhuang2023optimizing} uses slice-based decomposition with non-uniform, point-to-point sender–receiver mappings, while NTP~\cite{arfeen2025nonuniform} reconstructs temporary uniform layouts through point-to-point transfers before invoking standard collectives. In contrast,~\system employs LCM-based chunk partitioning and multi-ring collectives, enabling balanced synchronization across heterogeneous TP/DP/PP configurations without relying on sequential resharding phases.

\myparab{Full-stack simulators for distributed training.}
Full-stack AI training simulators such as ASTRA-sim~\cite{won2023astra}, SimAI~\cite{wang2025simai}, Multiverse~\cite{gui2025accelerating}, Phantora~\cite{qin2025phantora}, and ATLAHS~\cite{shen2025atlahs} model both system and network behavior. ATLAHS targets general-purpose distributed applications rather than AI training, while ASTRA-sim and Meta’s Arcadia~\cite{Arcadia} rely on production traces and offer limited support for heterogeneous communication patterns. Phantora~\cite{qin2025phantora} executes unmodified ML frameworks in a distributed containerized environment but does not model compute or network heterogeneity.



\myparab{Analytical full-stack simulators.}
Analytical simulators such as Echo~\cite{feng2024echo}, Neusight~\cite{lee2025forecasting}, and vTrain~\cite{bang2024vtrain} prioritize simulation speed over fidelity. However, these approaches trade modeling depth for speed and do not capture complex heterogeneous collective behavior.


\section{Conclusion} \label{sec:discussion}
\system enables faithful simulation of heterogeneous hybrid-parallel training through device groups, sweep-line DP formation, LCM-based synchronization, and gradient resharding.
\system’s extensible abstractions support emerging tensor-resharding and pipeline-parallel techniques, while dual NS-3 and htsim backends provide a flexible trade-off between network fidelity and simulation scalability.

We position~\system as a foundational step towards a unified evaluation platform for heterogeneous AI training. As models, hardware, and parallelization strategies co-evolve,~\system equips the community to systematically study their interactions and reason about future training infrastructures before deployment at scale.


\bibliographystyle{ACM-Reference-Format}
\bibliography{reference}

\appendix

\section{Input Specification Sample}
\label{sec:appendix:input_spec_example}

\begin{figure*}[t]
    \centering
    \includegraphics[width=1\linewidth]{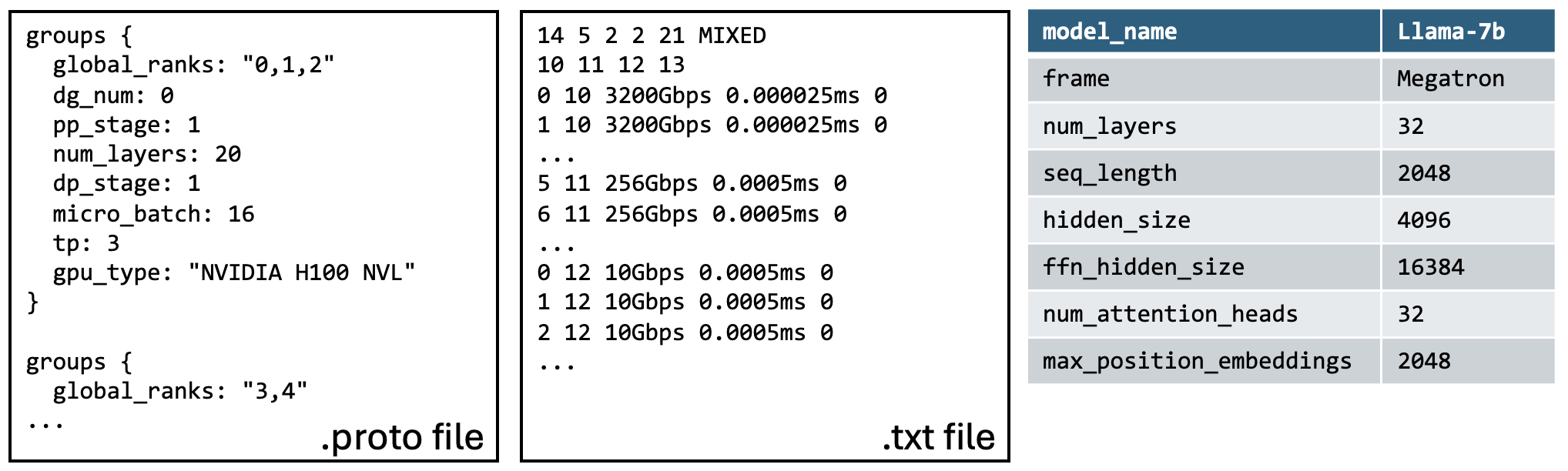}
    \caption{Input Specification for the example deployment configuration.}
    \label{fig:deployment_config}
\end{figure*}

The simulation framework mainly works with three input specifications which define our heterogeneous deployment configurations. Each configuration file captures distinct training setup, which enables us to give the simulator information about hybrid gpu clusters.

\myparab{Step 1. How are device groups and parallelism metadata defined?}\\
The first step initiates with creating a protocol buffer specification file that defines the device group configurations. A device group represents a collection of GPUs among which tensor parallelism must happen. As shown in the Figure~\ref{fig:deployment_config}, each device group entry specifies the global GPU ranks belonging to the group (e.g, ranks $\{0,1,2\}$ form a tensor parallel group of size 3) along with parallelism metadata. For instance, our given configuration with ranks $\{0,1,2\}$ at pipeline stage 1 and replica 1, execute the first set of transformer layers. This structured approach helps AICB to generate heterogeneous device group specific workloads.

\myparab{Step 2. Next! How can we model the heterogeneous network structure?}\\
This step involves creation of the topology file providing comprehensive network information for the simulation. From the Figure~\ref{fig:deployment_config} the file header specifies global topology parameters: total node count which consists of GPUs, NVLinks, and switches. Following the header, NVLink node identifier nodes are listed, then individual connection entries define the network structure. Adding our implementation the topology file captures heterogeneous link parameters reflecting real hardware differences, for example, H100 GPUs with 3200 Gbps NVLink bandwidth and 0.000025ms latency versus A100 GPUs with 256 Gbps NVLink bandwidth and 0.0005ms latency.

\myparab{Step 3. Finally! How are specific workloads generated for each group?}
The third steps the the model specifications and parameters which we provide AICB with to generate per device group workload files. Model parameters include hidden size (e.g., 4096 for LLaMA-7B), FFN hidden size (16384), number of layers (32), attention heads (32), vocabulary size (32000), and sequence length (2048). The workload generator which is built upon AICB (AI Collective Bechmark), processes this parameters along with the device group configurations to generate workload file per device group. Each of the workload file contains the training iteration structure: a header specifying the tensor parallel degree, pipeline parallel stages, gradient accumulation steps and embedding parameters. The header is then followed by work items for each layer including attention layers, MLP layers, and optimizer steps. Each work item encode with compute time(via GPU profiling), communication type (ALLREDUCE, ALLGATHER, REDUCESCATTER), communication size in bytes.

\section{Heterogeneity-aware Multi-Ring Construction Example}
\label{sec:appendix:multiring_example}

\begin{figure*}[t]
    \centering
    \includegraphics[width=\textwidth]{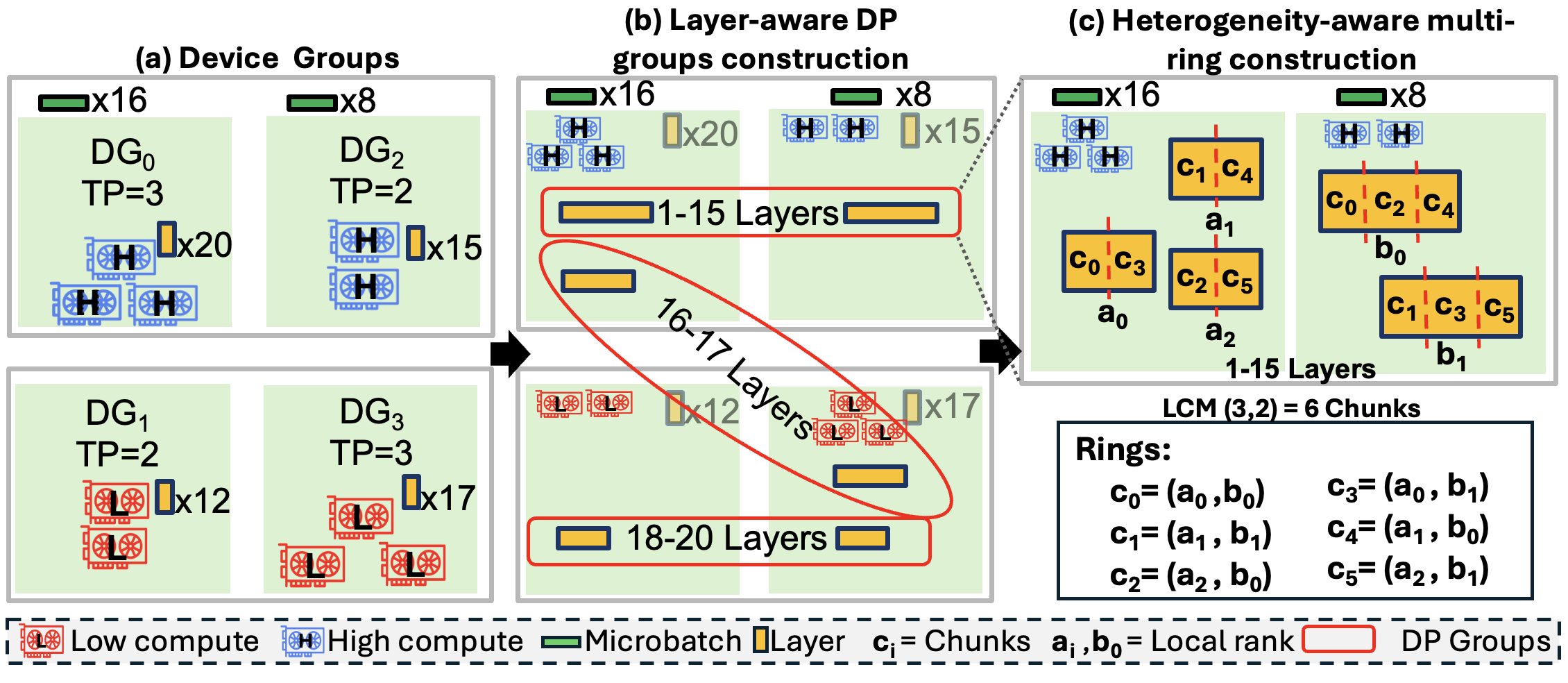}
    \caption{Heterogeneity-aware multi-ring LCM-based resharding in \system for non-uniform layer partitioning across high-compute (blue) and low-compute (red) device groups with layer-aware DP groups. }
    \label{fig:workflow-appendix}
\end{figure*}

Consider the deployment (\fref{fig:workflow-appendix}) of a model with 32 layers across four heterogeneous Device Groups with the following assigned layer ranges  $[s_i,e_i]$: $DG_0$ with layer range $[1,20]$ and ranks $\{0,1,2\}$, $DG_1$ with layer range $[21,32]$ and ranks $\{3,4\}$, $DG_2$ with layer range $[1,15]$ and ranks $\{5,6\}$ and, $DG_3$ with layer range $[16,32]$ and ranks $\{7,8,9\}$. This represents an asymmetric pipeline partitioning deployment and therefore, these Device Groups overlap on different portions of the model, implying that different subsets of ranks must synchronize gradients for different layer ranges. The sweep-line algorithm (Algorithm 1) essentially answers the question "\textbf{How do we find all these subsets?}" We illustrate the sweep-line algorithm using the above example.\\

\myparab{Step 1: Can we arrange the layer boundary points on a number line?}\\
Algorithm 1 begins by collecting the layer boundary points from all Device Groups. $For\ each$ Device Group with range $[s_i,e_i]$, the algorithm inserts $s_i$ and $(e_i+1)$ into the boundary set $\mathcal{P}$. We increment the end boundary points ($e_i$) by one which ensures that adjacent segments are handled cleanly.\\
For this example, $\mathcal{P}$ = $\{1,21,21,33,1,16,16,33\}$.

\myparab{Step 2: Looks a bit messy. Let's sort and clean!}\\
The boundary set $\mathcal{P}$ is then sorted and de-duplicated, which yields: $\mathcal{P}_{\text{unique}}$ = $\{1,16,21,33\}$. Let $p_i$ be the $i^\text{th}$ boundary point in $\mathcal{P}_{\text{unique}}$. From these boundary points, we induce the contiguous layer segments: \\
$[p_0,p_1-1] = \{1,15\}$\\ $[p_1,p_2-1] = \{16,20\}$\\ $[p_2,p_3-1] = \{21,32\}$ 

\myparab{Step 3: Cool! Now which Device Groups cover which segments?}\\
$For\ each$ segment $[seg_\text{start},seg_\text{end}]$, Algorithm 1 constructs a candidate set $C$ consisting of all the Device Groups whose layer ranges fully cover the segment.\\
\textbf{Segment $\textbf{[1,15]}$}: Both $DG_0\ [1,20]$ and $DG_2\ [1,15]$ fully cover this segment, so: $C$ = $\{DG_0,DG_2\}$.\\
\textbf{Segment $\textbf{[16,20]}$}: This segment is fully covered by $DG_0\ [1,20]$ and $DG_3\ [16,32]$, so: $C$ = $\{DG_0,DG_3\}.$\\
\textbf{Segment $\textbf{[21,32]}$}: This segment is covered by $DG_1\ [21,32]$ and $DG_3\ [16,32]$, so: $C$ = $\{DG_1,DG_3\}.$\\

\myparab{Step 3: Nice! Let's finally construct our DP synchronization groups}\\
$For\ each$ segment, if $|\mathcal{C}| \geq 2$, Algorithm 1 constructs a DP synchronization group $\mathcal{G}_{DP}$ by taking the union of ranks across all Device Groups in $C$.\\
\textbf{Segment $\textbf{[1,15]}$}: $\mathcal{G}_{DP}$ = $\{0,1,2\} \ \cup \ \{3,4\}$ = $\{0,1,2,3,4\}$ \\
\textbf{Segment $\textbf{[16,20]}$}: $\mathcal{G}_{DP}$ = $\{0,1,2\} \ \cup \ \{7,8,9\}$ = $\{0,1,2,7,8,9\}$ \\
\textbf{Segment $\textbf{[21,32]}$}: $\mathcal{G}_{DP}$ = $\{5,6\} \ \cup \ \{7,8,9\}$ = $\{5,6,7,8,9\}$ \\

Voila! There we have our DP groups. Each $\mathcal{G}_{DP}$ is associated with its corresponding layer segment.\\

We illustrate Algorithm 2 using the following. Consider a DP synchronization group which is responsible for the layer range $[1,15]$. This group consists of two Device Groups: $DG_0$ with tensor parallelism degree TP=3 and ranks $\{0,1,2\}$ and $DG_2$ with $TP=2$ and ranks $\{3,4\}$.\\

\myparab {Step 1: Which Device Groups participate?} \\
For the layer range $[1,15]$, the sweep-line algorithm has already determined that $DG_0$ and $DG_2$ will be synchronizing gradients. Algorithm 2 therefore takes these 2 Device Groups as the input DP group.

\myparab {Step 2: Wait! But how many chunks and rings?}\\
Since the participating Device Groups use different tensor parallelism degrees, their gradient tensors are partitioned differently. Algorithm 2 resolves this mismatch by computing the least common multiple ($LCM$) of the TP degrees of the participating Device Groups. In this example, $lcm(3,2)=6$. This value defines the number of gradient chunks to be communicated across the participating Device Groups inside the DP group and, hence, the number of communication rings required for the correct synchronization. Algorithm 2 conceptually represents each gradient as 6 equal sized chunks. Ranks in $DG_0$ collectively cover these chunks with 2 chunks per rank, and ranks in $DG_2$ cover 6 chunks with 3 chunks per rank. This representation helps in abstracting the physical tensor layouts while ensuring compatibility in synchronization across Device Groups. \\
\myparab{Step 3: Okay! Now, which chunk goes to which rank?}\\
Algorithm 2 assigns chunks to ranks in an interleaved/round-robin fashion based on each rank's local index within its Device Group. In this example, chunks are indexed from 0 to 5, and for each chunk $c$, the algorithm selects the participating ranks in each Device Group as 
$\text{local\_rank} = c \bmod \mathrm{TP}$. For $DG_0$, this maps chunks $\{0,3\}$ to local rank 0, $\{1,4\}$ to local rank 1, and $\{2,5\}$ to local rank 2. For $DG_2$, chunks $\{0,2,4\}$ map to local rank 0 and chunks $\{1,3,5\}$ to local rank 1. \\
\myparab{Step 4: Where are my rings?}\\
With the above mapping, Algorithm 2 constructs one communication ring per chunk. Each ring connects exactly one rank from $DG_0$ and one rank from $DG_2$ that owns the same chunk, ensuring that all participants in a ring operate on identically sized tensors. The interleaved assignment helps in distributing all communication evenly across the ranks and assign multiple chunks to a single rank.

\section{Heterogeneity-aware Chunk Partitioning Example}
We illustrate Algorithm 3 using the example of the DP synchronization group above. Consider a DP synchronization group $\mathcal{G_{DP}}$ responsible for synchronizing gradients for the layer range $[1,50]$. As per the sweep-line algorithm (Algorithm 1), this group consists of two Device Groups:\\\\
$DG_0$: TP degree $t_0 = 3$, ranks $\{0,1,2\}$\\
$DG_2$: TP degree $t_1 = 3$, ranks $\{3,4\}$\\\\

Let the total gradient communication volume for this DP group be $d=60\ MB$. Since the two Device Groups have different TP degrees, their gradients are initially partitioned into incompatible tensor shapes and hence, cannot be directly aggregated using a standard collective like AllReduce. The goal of Algorithm 3 is to resolve this mismatch into a common number of finer grained equal sized chunks.\\

\myparab{Step 1: How many chunks should be there in my $\mathcal{G_{DP}}$?}\\
Algorithm 3 starts by computing the least common multiple of the participating TP degrees: $L=lcm(t_0,t_1)=lcm(3,2)=6$. $L$ represents the total number of local gradient chunks used for synchronization with respect to $\mathcal{G_{DP}}$. This ensures tensor compatibility across heterogeneous TP configurations.

\myparab{Step 2: Umm...What is my per-rank gradient volume per Device Group?}\\
$For \ each$ Device Group $DG_i  \in  \mathcal{G_{DP}}$, Algorithm 3 computes the gradient volume handled by a single rank in $DG_i$: \\\\
For $DG_0$: $\text{data\_per\_rank}_{DG_{0}}=\dfrac{d}{t_0}=\dfrac{60 \ MB}{3}=20 \ MB$\\\\
For $DG_2$: $\text{data\_per\_rank}_{DG_{2}}=\dfrac{d}{t_2}=\dfrac{60 \ MB}{2}=30 \ MB$\\\\
This ensures that under tensor parallelism, each rank initially owns an equal fraction of the gradient within its Device Group.

\myparab{Step 3: But $(\text{data\_per\_rank}_{DG_{0}} \neq \text{data\_per\_rank}_{DG_{2}})$. So how do we make the chunk sizes equal?}\\
Next, Algorithm 3 determines how many chunks each rank in a Device Group will participate in: $\text{chunk\_multiplier}_{DG_i}=\dfrac{L}{t_i}$\\\\
For $DG_0$: $\text{chunk\_multiplier}_{DG_{0}}=\dfrac{L}{t_0}=\dfrac{6}{2}=3$\\\\
For $DG_2$: $\text{chunk\_multiplier}_{DG_{2}}=\dfrac{L}{t_2}=\dfrac{6}{3}=2$\\\\
This means that each rank in $DG_0$ contributes to 2 chunks and each rank in $DG_2$ contributes to 3 ranks. From this information, Algorithm 3 computes the volume of data contributed by each rank to each chunk:\\\\ $\text{data\_per\_chunk\_per\_rank}=\dfrac{\text{data\_per\_rank}_{DG_i}}{\text{chunk\_multiplier}_{DG_i}}$.\\\\
For $DG_0$: $\text{data\_per\_chunk\_per\_rank}=\dfrac{20 \ MB}{2}=10 \ MB$\\\\
For $DG_2$: $\text{data\_per\_chunk\_per\_rank}=\dfrac{30 \ MB}{3}=10 \ MB$\\\\

Thus, regardless of the uneven TP degrees, \textbf{every participating rank contributes exactly $\textbf{10 MB}$ to each chunk it owns.}

\label{sec:appendix:chunk_partitioning_example}

\section{Additional Evaluation Results}
\label{sec:evaluation-extended}



\begin{figure*}[t]
    \centering
    \captionsetup{font=small}
    
    \begin{minipage}[t]{0.28\linewidth}
        \centering
        \includegraphics[width=\linewidth]{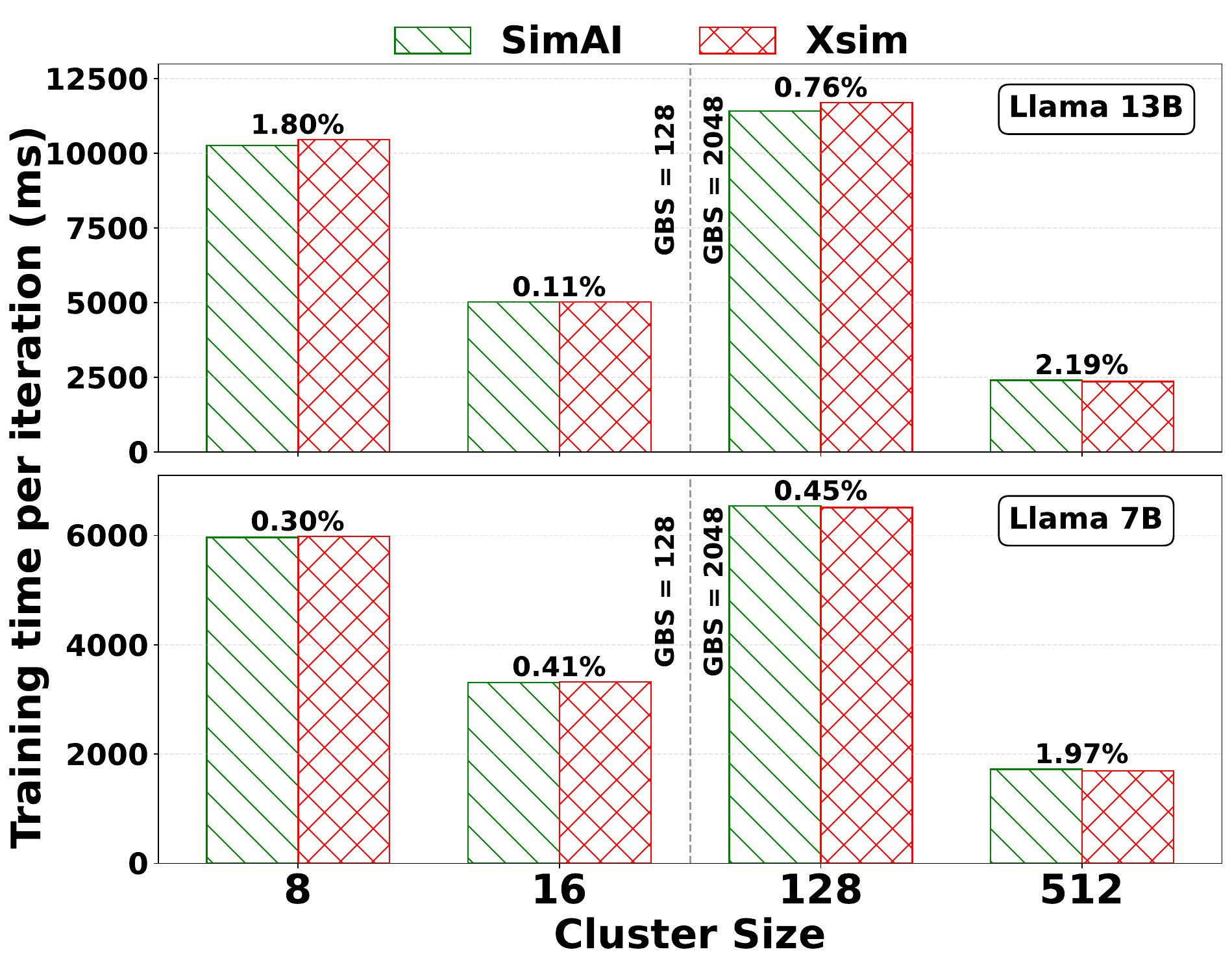}
        \caption{Training time per iteration for Llama 7B and 13B across homogeneous cluster sizes shows that \system closely matches SimAI, with a relative error of 0.1–2.2\%.}
        \label{fig:qus1-plot1}
    \end{minipage}
    \hfill
    \begin{minipage}[t]{0.32\linewidth}
        \centering
        \includegraphics[width=\linewidth]{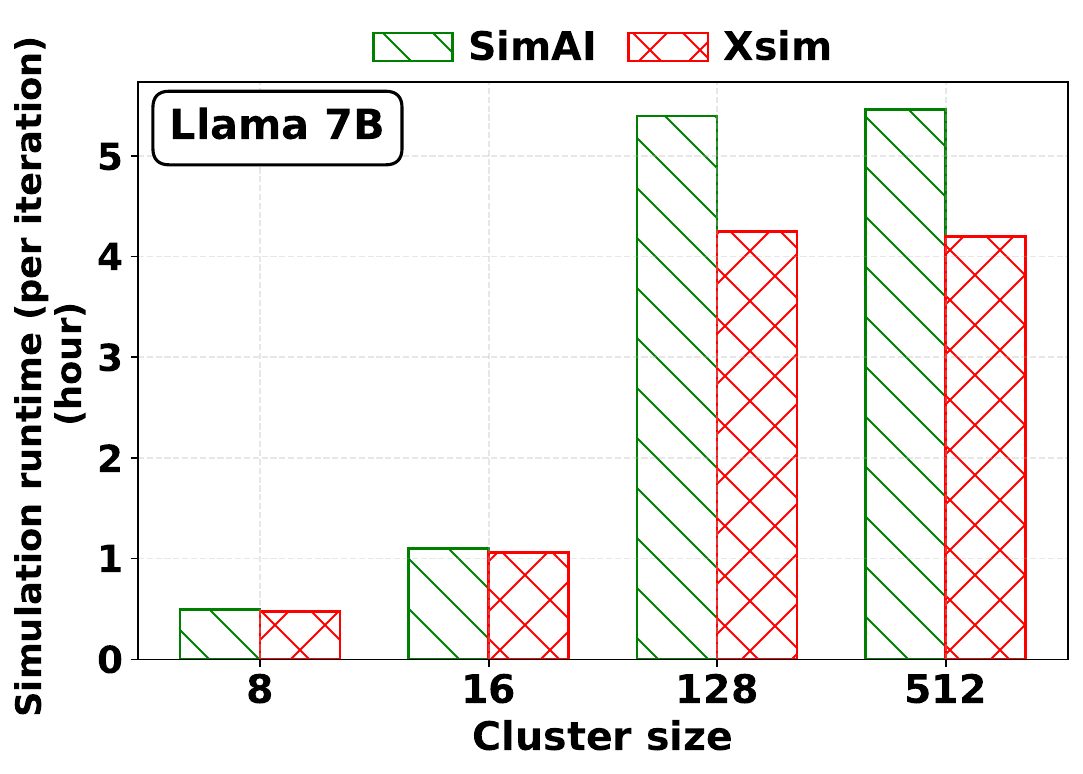}
        \caption{Simulation runtime per iteration for Llama 7B across homogeneous cluster sizes shows that \system achieves lower runtime at large scales (128–512 GPUs) due to LCM-based chunking and reduced communication overhead, with negligible differences at small scales.}
        \label{fig:qus1-plot2}
    \end{minipage}
    \hfill
    \begin{minipage}[t]{0.32\linewidth}
        \centering
        \includegraphics[width=\linewidth]{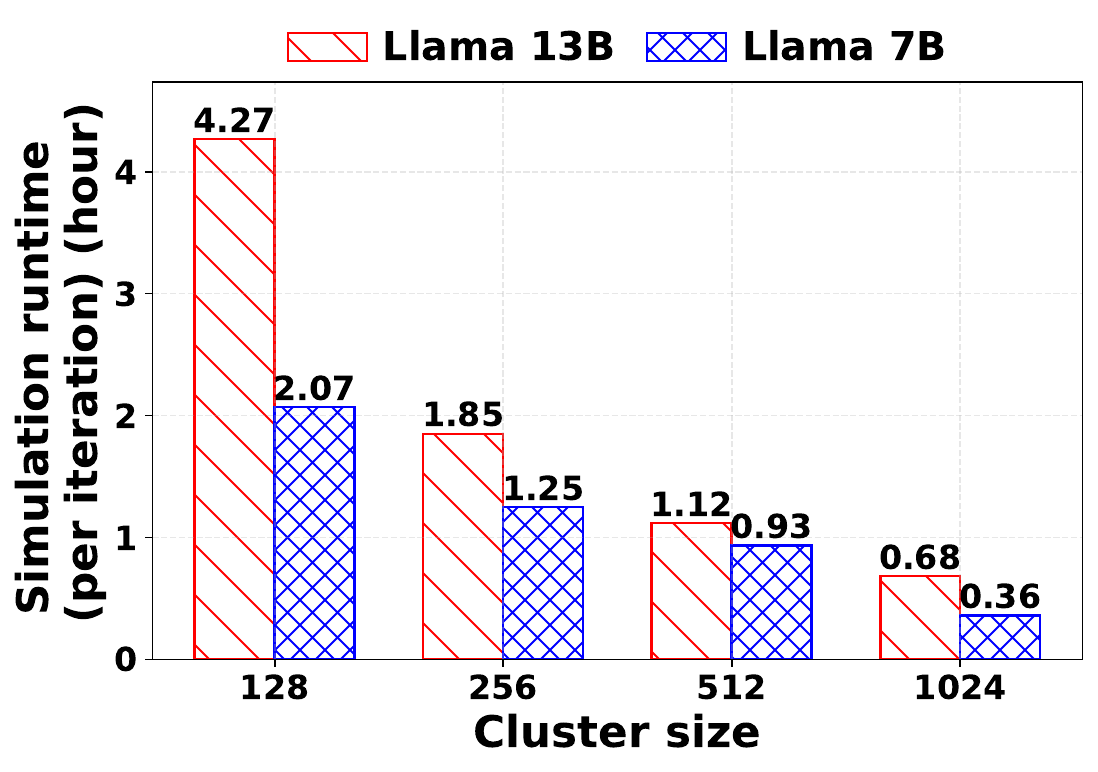}
        \caption{Simulation runtime per iteration versus heterogeneous cluster size for Llama 7B and 13B shows that \system scales efficiently, achieving up to 5.75× speedup for 7B and 6.25× for 13B as cluster size increases.}
        \label{fig:qus4-plot}
    \end{minipage}

\end{figure*}

\myparab{(8.a) Comparing~\system's prediction accuracy with SOTA training simulator, SimAI, for homogeneous AI clusters.}
Given that SimAI~\cite{wang2025simai} has demonstrated high predictive accuracy for homogeneous clusters using real-world experiments, we use it as a reference to validate the correctness of~\system’s modeling pipeline.
\fref{fig:qus1-plot1} reports the predicted training time (per iteration) for Llama 7B and Llama 13B across varying homogeneous cluster sizes. \system closely matches SimAI across all configurations, with relative errors between 0.1–2.2\%, validating the correctness of \system’s end-to-end modeling pipeline for homogeneous clusters.  

\myparab{(8.b) Comparing~\system's simulation time with SOTA training simulator, SimAI, for homogeneous AI clusters.}
\fref{fig:qus1-plot2} demonstrates the wall-clock time of the simulator, i.e., the simulation runtime for Llama 7B over varying cluster sizes. 
At large cluster scales (i.e., 128 and 512 GPUs),~\system achieves lower per-iteration simulation runtime than SimAI. This improvement stems from \system’s LCM-based chunking and chunk partitioning, which limit chunk fragmentation and reduce the number of simulated communication events. In contrast, SimAI’s mockNCCL implementation increases the number of chunks as the cluster size grows, even for small models, leading to higher event-scheduling and synchronization overhead during simulation. As a result,~\system incurs lower wall-clock simulation time at scale. These differences are negligible at small cluster sizes but become pronounced as collective communication dominates execution at higher degrees of parallelism.




\begin{figure*}[t]
    \centering
    \captionsetup{font=small}

    \begin{minipage}[t]{0.48\linewidth}
        \centering
        \includegraphics[width=\linewidth]{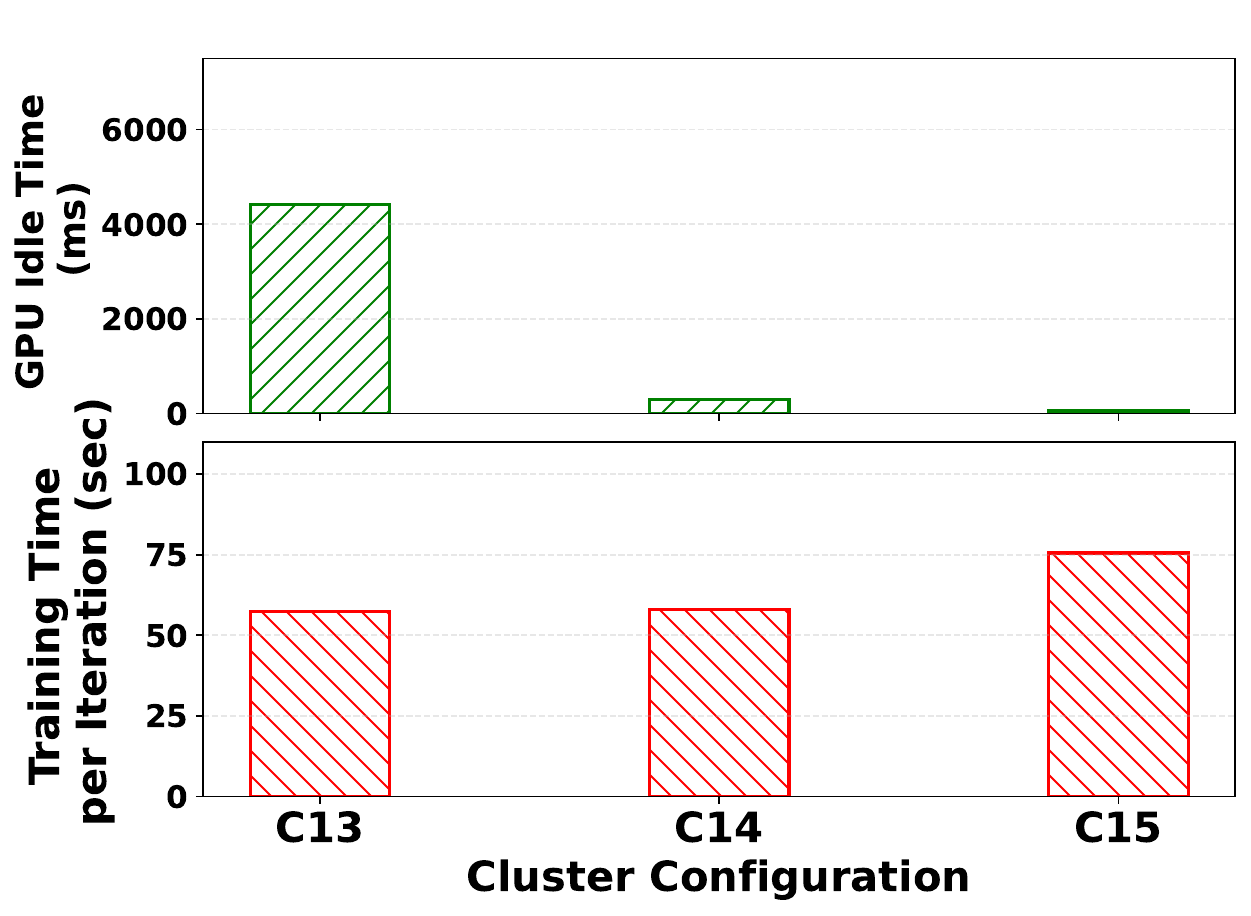}
        \caption{GPU idle time and training time per iteration across cluster configurations ((C13–C15) Table~\ref{tab:config-model-cluster}), highlighting imbalance and utilization differences.}
        \label{fig:qus6-plot1}
    \end{minipage}
    \hfill
    \begin{minipage}[t]{0.48\linewidth}
        \centering
        \includegraphics[width=\linewidth]{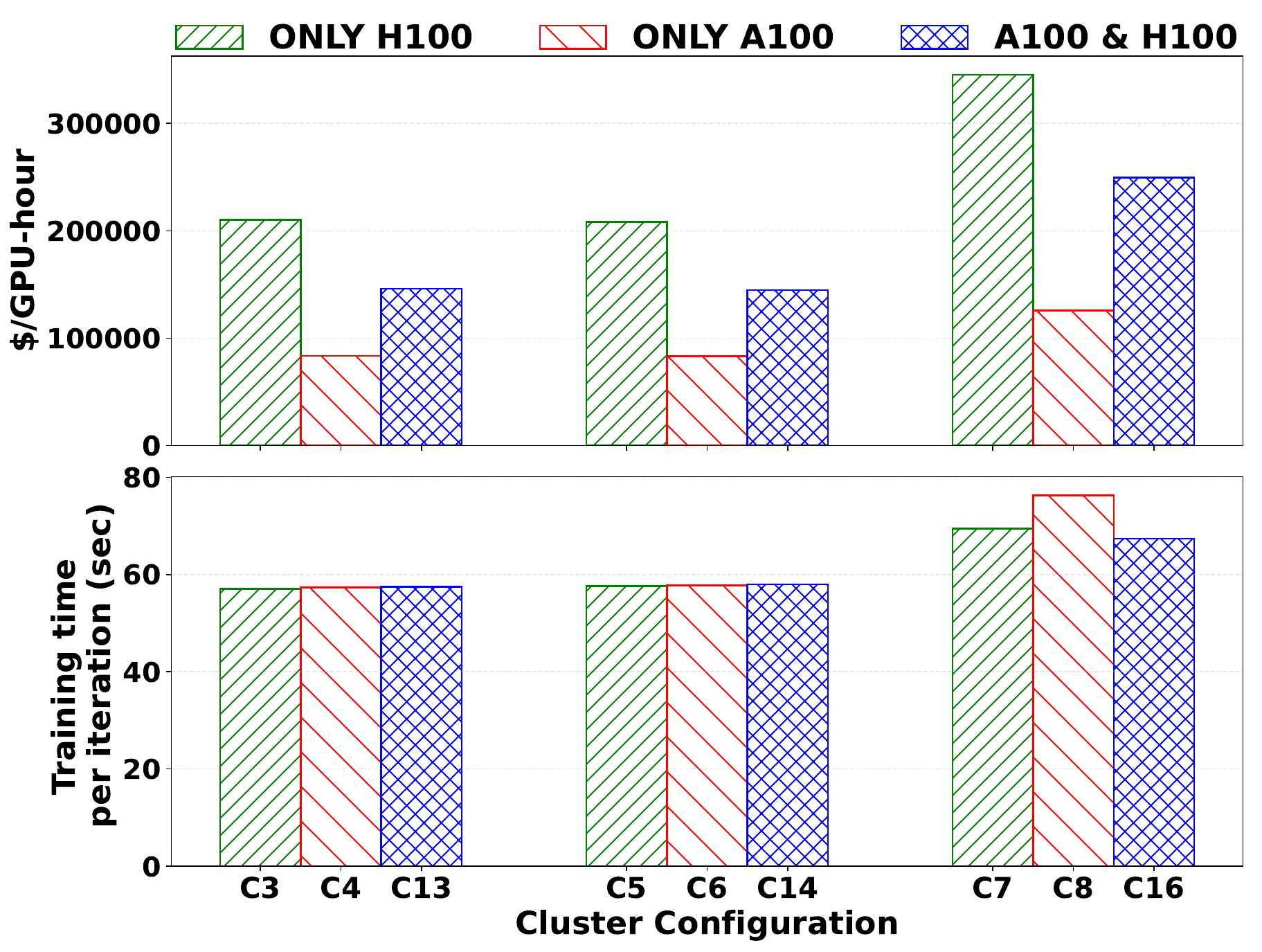}
        \caption{Cost per GPU-hour and training time per iteration for Llama 7B across homogeneous and heterogeneous clusters.}
        \label{fig:qus6-plot2}
    \end{minipage}

\end{figure*}

\myparab{(9) \system's scalability analysis. }
We evaluate~\system's scalability across model sizes (Llama 7B and Llama 13B) and heterogeneous cluster configurations, i.e., $C15$, with varying cluster size (see~\fref{fig:qus4-plot}).
As cluster size increases, the simulation runtime decreases substantially, achieving up to $5.75\times$ speedup for Llama 7B and $6.25\times$ for Llama 13B models. For any fixed cluster, the Llama 13B model incurs higher runtime due to increased computation and communication. 
The performance gap narrows at mid-scale configurations due to fewer simulated network events per iteration. As cluster size increases, collective communication is partitioned across more ranks and rings, resulting in smaller per-flow data sizes and more regular traffic patterns.
These results demonstrate that~\system scales efficiently with both model and cluster size in heterogeneous environments.

\myparab{(10.a) \system's utility: How can {GPU idle time} or straggler analysis help deployers?}
\fref{fig:qus6-plot1} shows the model training time, and the GPU idle time (i.e., the Straggler waiting time)  across {\em three} heterogeneous configurations, $C13$ (DP), $C14$ (TP+DP), and $C15$ (TP+PP) (see Table~\ref{tab:config-model-cluster}).
We observe that the heterogeneous DP configuration exhibits a waiting time of $4.42~sec$, while switching to a TP+DP configuration reduces waiting time to $0.294~sec$, and further reduces to $0.07~sec$ under PP+TP, demonstrating how workload partitioning mitigates synchronization stalls.
More interestingly, we see that the training time for $C13$ and $C14$ is similar, but the GPUs with $C13$ are underutilized. One may optimize the algorithm or use GPU time for alternative tasks, such as checkpointing.



\myparab{(10.b) \system's utility: Total Cost of Ownership (TCO)}
Across homogeneous and heterogeneous cluster designs using Llama 7B (see~\fref{fig:qus6-plot2}), we observe that heterogeneous clusters can achieve near-identical or superior performance while significantly reducing cost. 
For instance, in the 8-GPU data-parallel configuration ($C3$, $C4$, $C13$), a mixed $4\times$H100 + $4\times$A100 cluster achieves a completion time of 57.48\,s, comparable to the $8\times$H100 cluster (57.06\,s), yet reduces TCO from $210K$ \$/GPU-hour to $146K$ \$/GPU-hour, yielding a $\sim$30\% cost-efficiency improvement. Similarly, compared to an $8\times$A100 cluster (TCO = $83K$ \$/GPU-hour), the heterogeneous configuration offers higher performance headroom at moderate additional cost. 
Interestingly, for the configurations $C9$, $C10$, and $C16$, we observe that the heterogeneity solution performs best and has a lower cost than the $8\times$H100 cluster. This could be because the non-uniform hybrid parallelism led to better resource utilization.

\section{\system's LCM-based tensor resharding: Bounds and Tradeoffs} 
\label{sec:appendix:lcm_bounds}
Since practical deployments use TP degree, $t_i \leq 8$, the maximum possible synchronization granularity is bounded by
$\mathrm{LCM}(t_1,\ldots,t_k) \leq 8 \times 7 \times 5 \times 3 = 840$,
which occurs when the TP degrees collectively contain all prime factors up to 8. Thus, the worst-case LCM remains bounded by 840 regardless of the number of participating device groups ($k \geq 4$).

For an AllReduce with $k$ participants and message size $c$, the communication cost is approximately
$T_{\text{ring}} \approx 2(k-1)\left(\alpha + \frac{c}{kB}\right)$,
for a ring algorithm, and
$T_{\text{tree}} \approx 2\log_2(k)\left(\alpha + \frac{c}{B}\right)$,
for a binary-tree algorithm, where $\alpha$ is the per-message latency and $B$ is the link bandwidth.

In large-scale deployments ($>16K$ GPUs), hierarchical collectives typically use local groups with $t_i \leq 8$, resulting in a global DP group of roughly $16{,}000/8 \approx 2{,}000$ participants. Further sharding (e.g., FSDP~\cite{zhao2023pytorch}) commonly operates on smaller subgroups (e.g., 128 ranks). For a 4\,GB gradient synchronization (approximately the gradient size of Llama-2 70B), the per-rank chunk size is only
$\frac{4\ \mathrm{GB}}{128} \approx 32\ \mathrm{MB}$,
indicating that the bounded LCM and chunk-partitioning overheads remain practical even at large scales.


\section{Extended Related Work}
\label{sec:related-extended}

\myparab{High-speed simulators.}
Nüwa~\cite{li2025nuwa} accelerates AI network simulation by reducing routing construction overhead via formula-based routing. Multiverse~\cite{gui2025accelerating} achieves speedups using a GPU-accelerated design. FPGA-based simulators such as Miniature~\cite{qian2025miniature} achieve orders-of-magnitude acceleration by mapping homogeneous network components to specialized hardware, scaling to 65K nodes. 
Recent packet-level simulators such as Wormhole~\cite{long2026supercharging} and HyGra~\cite{wang2026hygra} accelerate LLM network simulation through steady-state detection, state reuse, network partitioning, and adaptive switching between packet- and flow-level models. While these techniques significantly reduce simulation time, they assume homogeneous infrastructure and fixed communication semantics. \system complements these efforts by providing heterogeneity-aware full-stack simulation, while supporting htsim as a scalable software backend for faster network simulation without specialized hardware, albeit with lower acceleration than hardware-assisted approaches.

\end{document}